\documentclass[runningheads,orivec]{llncs}
\usepackage[T1]{fontenc}
\usepackage{graphicx}
\usepackage[labelformat=simple]{subcaption}
\usepackage{xcolor}
\usepackage[T1]{fontenc}
\usepackage[utf8]{inputenc}
\usepackage{listings}
\usepackage{array}

\definecolor{hypcolor}{rgb}{0.85, 0.48, 0.0}
\lstdefinestyle{leanHH}{
  language=lean,
  moredelim=**[is][\color{hypcolor}]{|!}{!|},
}
\usepackage{amssymb}
\usepackage[misc]{ifsym}
\usepackage{listings}
\usepackage{stys/verus}
\usepackage{stys/smtlib}

\usepackage{cite}
\usepackage{color}       %
\usepackage{pifont}      %
\usepackage{booktabs} %

\lstdefinestyle{DefaultStyle}
{
  basicstyle=\scriptsize\fontfamily{pxtt}\linespread{0.8}\selectfont,
  breaklines=true,
  tabsize=1,
  breakindent=2em,
  literate={\ \ }{{\ }}1
}

\newcommand{\vcode}[1]{\sloppy\inlineverus{#1}\xspace}
\newcommand{\smtcode}[1]{\sloppy\lstinline[language=smtlib, basicstyle=\ttfamily\normalsize]{#1}\xspace}

\makeatletter
\def\sectionautorefnamesymbol{Sec. \@gobble}

\makeatother

\newif\ifsubmit    %
\submittrue

\ifsubmit
  \newcommand{\comment}[1]{}
  \newcommand{\todo}[1]{}
  
\else
  \newcommand{\todo}[1]{\ding{110}\ding{43}\textcolor{red}{TODO: #1}}
  \newcommand{\comment}[1]{\ding{110}\ding{43}\textcolor{red}{#1}}
  
\fi

\definecolor{keywordcolor}{rgb}{0.0, 0.0, 0.0}   %
\definecolor{tacticcolor}{rgb}{0.0, 0.3, 0.8}    %
\definecolor{commentcolor}{rgb}{0.4, 0.4, 0.4}   %
\definecolor{symbolcolor}{rgb}{0.0, 0.0, 0.0}    %
\definecolor{sortcolor}{rgb}{0.0, 0.0, 0.0}      %
\definecolor{attributecolor}{rgb}{0.7, 0.1, 0.1} %

\usepackage{fancybox}

\usepackage{amsmath}
\usepackage{algorithm2e}
\SetKwProg{Fn}{Function}{}{end}
\SetKwSwitch{Switch}{Case}{Other}{match}{}{case}{otherwise}{}{}
\SetKwFor{For}{for}{do}{}
\SetKwInOut{Input}{In}
\SetKwInOut{Output}{Out}
\SetKw{Continue}{continue}
\SetKw{Break}{break}
\usepackage{tikz}
\usetikzlibrary{calc,trees,positioning,arrows.meta,chains,fit,backgrounds}
\usetikzlibrary{shapes.geometric,decorations.pathreplacing}
\usetikzlibrary{decorations.pathmorphing,shapes,matrix,shapes.symbols}
\usepackage{wrapfig}

\usepackage{enumitem}
\setlist{topsep=2pt, partopsep=0pt, itemsep=1pt, parsep=0pt}
\makeatletter
\renewcommand\section{\@startsection{section}{1}{\z@}%
                       {-12\p@ \@plus -2\p@ \@minus -2\p@}%
                       {6\p@ \@plus 2\p@ \@minus 2\p@}%
                       {\normalfont\large\bfseries\boldmath
                        \rightskip=\z@ \@plus 8em\pretolerance=10000 }}
\renewcommand\subsection{\@startsection{subsection}{2}{\z@}%
                       {-10\p@ \@plus -2\p@ \@minus -2\p@}%
                       {5\p@ \@plus 2\p@ \@minus 2\p@}%
                       {\normalfont\normalsize\bfseries\boldmath
                        \rightskip=\z@ \@plus 8em\pretolerance=10000 }}
\renewcommand\subsubsection{\@startsection{subsubsection}{3}{\z@}%
                       {-8\p@ \@plus -2\p@ \@minus -2\p@}%
                       {-0.5em \@plus -0.22em \@minus -0.1em}%
                       {\normalfont\normalsize\bfseries\boldmath}}
\renewcommand\paragraph{\@startsection{paragraph}{4}{\z@}%
                       {-6\p@ \@plus -2\p@ \@minus -2\p@}%
                       {-0.5em \@plus -0.22em \@minus -0.1em}%
                       {\normalfont\normalsize\itshape}}
\makeatother
\AtBeginDocument{%
  \setlength\abovedisplayskip{4pt plus 2pt minus 2pt}%
  \setlength\belowdisplayskip{4pt plus 2pt minus 2pt}%
  \setlength\abovedisplayshortskip{0pt plus 2pt}%
  \setlength\belowdisplayshortskip{2pt plus 2pt minus 1pt}}

\usepackage[hypertexnames = false]{hyperref}

\hypersetup{
  colorlinks=true,
  linkcolor=blue,
  urlcolor=blue,
  citecolor=blue
}
\usepackage{orcidlink}
\usepackage[numbers]{natbib}

\newcommand{\toolname}{Sundance\xspace}

\newcommand{\fstar}{F${}^\star$\xspace}
\newcommand{\cadical}{CaDiCaL\xspace}

\newcommand{\mariposa}{Mariposa\xspace}

\newcolumntype{K}[1]{>{\centering\arraybackslash}p{#1}}
\DeclareRobustCommand{\refappendix}[1]{%
  \ifdefined\cameraReady
    Appendix~\ref*{#1} of \cite{ThisPaperOnArxiv}%
  \else
    Appendix~\ref{#1}%
  \fi
}

\DeclareRobustCommand{\maybeappendix}[1]{%
  \ifdefined\cameraReady
    \refstepcounter{section}
  \else
    \section{#1}
  \fi
}

\begin{document}

\title{\toolname: A Simple SMT Solver for Program Verification}

\author{
  Amar Shah\inst{1,2},
  Benjamin F.~Jones\inst{2},
  Can Cebeci\inst{3},
  Jason Z.~S. Hu\inst{2},
  Lee A. Barnett\inst{2},
  R\'{e}mi Delmas\inst{2},
  Marijn Heule\inst{1,2},
  Bryan Parno\inst{1,2},
  Bruno Dutertre\inst{2}
}

\institute{
  Carnegie Mellon University, Pittsburgh, USA, \and
  Amazon Web Services, Seattle, USA \and
  EPFL, Lausanne, Switzerland
}

\authorrunning{A. Shah et al.}

\maketitle              %

\begin{abstract}
We present \toolname, an SMT solver for program verification written in about
13,600 lines of Rust. \toolname combines the \cadical SAT solver with congruence
closure, linear arithmetic, algebraic datatypes, and pattern-based quantifier
instantiation. \toolname produces eDRAT proofs for every \texttt{unsat} answer.
On 21,531 queries from Verus, Dafny, and \fstar, \toolname is comparable to
existing solvers, with 26--38$\times$ fewer lines of code than Z3 and cvc5. We
further show that its components can be swapped for Z3's arithmetic solver,
MiniSat, and the Semper e-graph. We have open sourced \toolname.
\end{abstract}

\section{Introduction}
\label{sec:introduction}

Auto-active program verification tools such as Dafny~\cite{dafny},
Verus~\cite{verus-ghost}, and \fstar~\cite{fstar} are used to verify
large-scale systems in industry and
academia~\cite{amazon-cloudscale,galapagos,aws-encryption-sdk-dafny,nitro-isolation-engine,everparse3d,ironfleet,verus-sys,evercrypt}. Program
verification has become widespread in part because of automation
provided by off-the-shelf Satisfiability Modulo Theories (SMT)
solvers. Verification tools translate program properties into
first-order logic and call a solver, such as Z3~\cite{z3},
cvc5~\cite{cvc5}, or Alt-Ergo~\cite{altergo} in the back-end. This
allows verification engineers to focus on high-level program
properties and ignore low-level proof details.

Z3 and cvc5, the primary SMT solvers used in program verification,
support many logical fragments and theories. They are employed for a
wide variety of tasks: mathematics~\cite{satcas},
fuzzing~\cite{smtfuzzing}, model-checking~\cite{smtmodelchecking}, and
many others. This requires a broad set of capabilities, complicated
heuristics, and performance tuned for many different use cases. This
leads to complex, large code bases: Z3 and cvc5 are each several
hundred thousand lines of C++ code (519,367 lines for Z3 and 358,294 for
cvc5)~\cite{z3,cvc5}.\footnote{Line counts exclude blank lines, comments, and
tests.}
This complexity costs verification users:
\begin{enumerate}
    \item \emph{Opacity.} It is difficult to understand why a solver succeeds or
    fails, so users resort to intricate encodings, obscure flag combinations,
    and bespoke debugging tools~\cite{axiom-profiler}.
    \item \emph{Trust issues.} Existing solvers are trusted by the
      verification tools, but they have been shown to contain
      bugs~\cite{yinyang}.
    \item \emph{Monolithic architectures.} Tightly coupled components make it difficult to add
    features or adopt new advances. For instance, there has been an explosion
    of research in e-graphs~\cite{egg, egglog}, but none of these new ideas 
    have made their way into mainstream SMT solvers.
\end{enumerate}

We believe it is time to revisit how we design SMT solvers.  Verifiers
use only a narrow fragment of SMT, and often disable general-purpose
features such as model-based quantifier instantiation. Meanwhile, the
building blocks of a solver have matured into reusable components:
modern SAT solvers such as \cadical~\cite{cadical} expose interfaces
for theory integration~\cite{ipasir-up}, e-graph libraries are widely
available~\cite{egg,egglog,semper}, and proof formats such as
eDRAT~\cite{edrat} make certification much cheaper than before.

To demonstrate this, we present \toolname, a simpler SMT solver
specialized for program verification. \toolname is written in 13,612
lines of Rust and consists of the \cadical SAT
solver~\cite{cadical}, a pattern-based quantifier instantiation
engine, a congruence closure module, and a linear arithmetic
module. These features are sufficient to support 91.3\% of the queries
in our held-out Verus benchmarks, 93.0\% of Dafny benchmarks, and
97.1\% of \fstar benchmarks. Verus and Dafny developers have asked for
a solver with exactly these
features~\cite{leino-personal-communication}.

\toolname has three key properties:
\begin{enumerate}
    \item \emph{Transparency}. \toolname is designed to be simple, with a
    clear architecture and few heuristics (\autoref{sec:egraph}--\autoref{sec:quantifiers}).
    \item \emph{Trustworthiness}. \toolname produces proofs in the
      eDRAT format, a clausal proof format inspired by SAT solvers,
      and every \texttt{unsat} result comes with a proof.
    \item \emph{Modularity}. Each module implements a Rust trait, so
    components can be swapped out: \toolname can use either \cadical or
    MiniSat as its SAT solver (\autoref{app:sat}), and either its own
    arithmetic solver or Z3's.
\end{enumerate}

We describe \toolname's architecture and main modules. We then evaluate \toolname
on 6,662 queries from Verus, including a held-out set of 4,086 queries not used
during development, and on 11,674 Dafny and 3,195 \fstar queries from the \mariposa
benchmark suite~\cite{mariposa}, comparing it with Z3, cvc5, and Alt-Ergo. Z3
without proofs solves nearly every query, which is expected: each query comes
from a project that was already verified with Z3, so Z3 solves it by
construction. Our evaluation shows that:

\begin{itemize}
  \item \textbf{\toolname is competitive on Verus} (\autoref{subsec:eval-verus}).
  On the held-out Verus set, it solves 86.7\% of queries within 10\,s, more than Z3
  with proofs (82.7\%) and Alt-Ergo (77.4\%) and close to cvc5 (90.5\%), and
  90.2\% within 60\,s, compared with 95.3\% for cvc5, 97.7\% for Z3 with proofs,
  and 82.8\% for Alt-Ergo.
  \item \textbf{\toolname generalizes to Dafny and \fstar}
  (\autoref{subsec:eval-mariposa}), even though it was not developed for them.
  On Dafny, it solves 73.7\% of queries, more than any other proof-producing
  solver and Alt-Ergo; on \fstar, it solves 56.7\%, between cvc5 (54.9\%) and
  Alt-Ergo (59.3\%).
  \item \textbf{\toolname's modules can be swapped}
  (\autoref{subsec:eval-modules}). \toolname also runs with Z3's arithmetic
  solver, the Semper e-graph~\cite{semper}, and the MiniSat SAT
  solver~\cite{minisat}. Each alternative regresses on every benchmark set, indicating that \toolname's default configuration is the
  best of those we tried.
\end{itemize}

\noindent We have \href{https://github.com/yaspar-org/Sundance-SMT}{open sourced \toolname}.

\section{\toolname's Design}
\label{sec:design}

We motivate the need for \toolname's components through a running
example from Verus and the SMT query it generates.
Proving this query requires Boolean case analysis, equality reasoning,
arithmetic, and quantifier instantiation.  We present \toolname's
architecture, and we describe each module and the part of the example it
is responsible for.

\subsection{Running Example}
\label{subsec:example}

\newcommand{\hlmark}[1]{\tikz[remember picture]\coordinate (#1);}
\colorlet{invhl}{yellow!70!orange}
\DeclareRobustCommand{\invhl}[1]{{\setlength{\fboxsep}{0.5pt}\colorbox{invhl!25}{\vphantom{Ag}#1}}}
\begin{figure}[t]
\centering
\begin{minipage}[t]{0.47\textwidth}
\begin{lstlisting}[language=Verus,style=VerusLineNos,firstnumber=1,
        basicstyle=\ttfamily\scriptsize,numberstyle=\tiny\color{gray},xleftmargin=10pt,numbersep=4pt]
fn upper_bound(a: &Vec<i32>) -> (ub: i32)
    ensures
        forall|i: int| 0 <= i < a.len() (*@\label{line:ub-ensures}@*)
            ==> a[i] <= ub,
{
    let mut ub = 0;
    for i in 0..a.len()
        invariant
            (*@\hlmark{invA}@*)forall|j: int| 0 <= j < i(*@\hlmark{invB}@*) (*@\label{line:ub-invariant}@*)
                ==> a[j] <= ub,(*@\hlmark{invC}@*)
    {
        if a[i] > ub {
            ub = a[i];
        }
    }
    return ub
}
\end{lstlisting}
\begin{tikzpicture}[remember picture, overlay]
  \fill[invhl, fill opacity=0.25, rounded corners=1pt]
    let \p1 = (invB), \p2 = (invC) in
    ([shift={(-1.5pt, 1.9ex)}]invA) rectangle ({max(\x1, \x2) + 1.5pt}, \y2 - 0.7ex);
\end{tikzpicture}%
        \end{minipage}%
\hfill
\begin{minipage}[t]{0.51\textwidth}
\begin{lstlisting}[ columns=fullflexible, language=smtlib,
        commentstyle=\color{green!50!black}, basicstyle=\ttfamily\scriptsize,
        morekeywords={}, numbers=left, numbersep=4pt,
        numberstyle=\tiny\color{gray}, xleftmargin=10pt, firstnumber=1,
        stepnumber=1 ]
(declare-sort Vec 0)
(declare-fun len (Vec) Int)
(declare-fun index (Vec Int) Int)

(declare-const a Vec)
(declare-const i Int)
(declare-const ub Int)

; Assume: i is in bounds
(assert (and (<= 0 i) (< i (len a)))) (*@\label{line:smt-indexing}@*)

; Loop invariant holds before iteration i
(assert (forall ((j Int)) (*@\label{line:smt-invariant1}@*)
  (! (=> (and (<= 0 j) (< j i))
         (<= (index a j) ub))
     :pattern ((index a j))))) (*@\label{line:smt-invariant2}@*)

; ub_new = (if a[i] > ub then a[i] else ub)
(declare-const ub_new Int)
(assert (= ub_new (*@\label{line:smt-ub-new}@*)
  (ite (> (index a i) ub) (index a i) ub))) (*@\label{line:smt-ub-new2}@*)

; Negated goal: invariant fails for i+1
(assert (not (forall ((j Int)) (*@\label{line:smt-goal-start}@*)
  (! (=> (and (<= 0 j) (< j (+ i 1)))
         (<= (index a j) ub_new))
     :pattern ((index a j)))))) (*@\label{line:smt-goal-end}@*)
\end{lstlisting}
\end{minipage}
\begin{minipage}[t]{0.47\textwidth}
\subcaption{Verus source.}\label{fig:verus-upper-bound} \end{minipage}%
\hfill
\begin{minipage}[t]{0.51\textwidth}
\subcaption{SMT-LIB query for the \invhl{invariant}.}\label{fig:smt-upper-bound}
\end{minipage}
\caption{Running example. (a)~A Verus function that computes an upper bound of a
vector, with its \invhl{loop invariant} highlighted. (b)~The simplified SMT-LIB
query that Verus generates to check that the \invhl{loop invariant} is preserved
by each iteration.}
\label{fig:running-example}
\end{figure}

\autoref{fig:verus-upper-bound} shows a function written in Verus that iterates
through a vector of integers, and maintains a running upper bound \vcode{ub}.
Its \vcode{ensures} clause (line~\ref{line:ub-ensures}) states that the returned
value is greater than or equal to every element of the vector. Verus cannot
prove this on its own, so the developer provides a loop invariant
(line~\ref{line:ub-invariant}): at the start of each iteration, \vcode{ub} is
greater than or equal to every element seen so far.  Verus must then show that
(1)~the invariant holds on entry to the loop, (2)~it is preserved by each
iteration, and (3)~it implies the \vcode{ensures} clause on exit. Each of these
obligations becomes a separate SMT query.

\autoref{fig:smt-upper-bound} shows a simplified version of the SMT
query for checking that the loop invariant is preserved. The original
is about 6,000 lines long, since Verus includes a large prelude of axioms for vectors and other
standard constructs~\cite{verus-ghost}.

The query models vectors with an uninterpreted sort \smtcode{Vec} and
uninterpreted functions \smtcode{len} and \smtcode{index}. The constants
\smtcode{i} and \smtcode{ub} are the loop index and bound at the start of an
arbitrary iteration. Line~\ref{line:smt-indexing} assumes the loop index is in
bounds, and lines~\ref{line:smt-invariant1}--\ref{line:smt-invariant2} assume
the invariant holds at the start of the iteration.
Lines~\ref{line:smt-ub-new}--\ref{line:smt-ub-new2} define \smtcode{ub_new}, the
bound after the loop body. Finally,
lines~\ref{line:smt-goal-start}--\ref{line:smt-goal-end} assert the
\emph{negation} of the invariant for the next iteration. The obligation holds if
and only if the query is unsatisfiable.

The query is unsatisfiable: the negated goal
(lines~\ref{line:smt-goal-start}--\ref{line:smt-goal-end}) states that
the invariant fails for $i+1$. There is then some index $j_0$ such that $0
\le j_0 < i+1$ and $\mathit{index}(a, j_0) > \mathit{ub\_new}$. By its
definition (lines~\ref{line:smt-ub-new}--\ref{line:smt-ub-new2}),
$\mathit{ub\_new}$ is the larger of $\mathit{ub}$ and
$\mathit{index}(a, i)$.  If $j_0 < i$, the invariant
(lines~\ref{line:smt-invariant1}--\ref{line:smt-invariant2}) applies
to $j_0$, so $\mathit{index}(a, j_0) \le \mathit{ub} \le
\mathit{ub\_new}$. Otherwise, since $j_0$ is an integer with $i \le
j_0 < i+1$, we have $j_0 = i$, and so $\mathit{index}(a, j_0) =
\mathit{index}(a, i) \le \mathit{ub\_new}$. Both cases contradict
$\mathit{index}(a, j_0) > \mathit{ub\_new}$.

Each step of this argument is carried out by a different part of \toolname:
naming $j_0$ is skolemization, applying the invariant to $j_0$ is quantifier
instantiation, the case split is made by the SAT solver, $j_0 = i$ requires
integer arithmetic, and $\mathit{index}(a, j_0) = \mathit{index}(a, i)$ follows
by congruence.

\subsection{Architecture}
\label{subsec:overview}

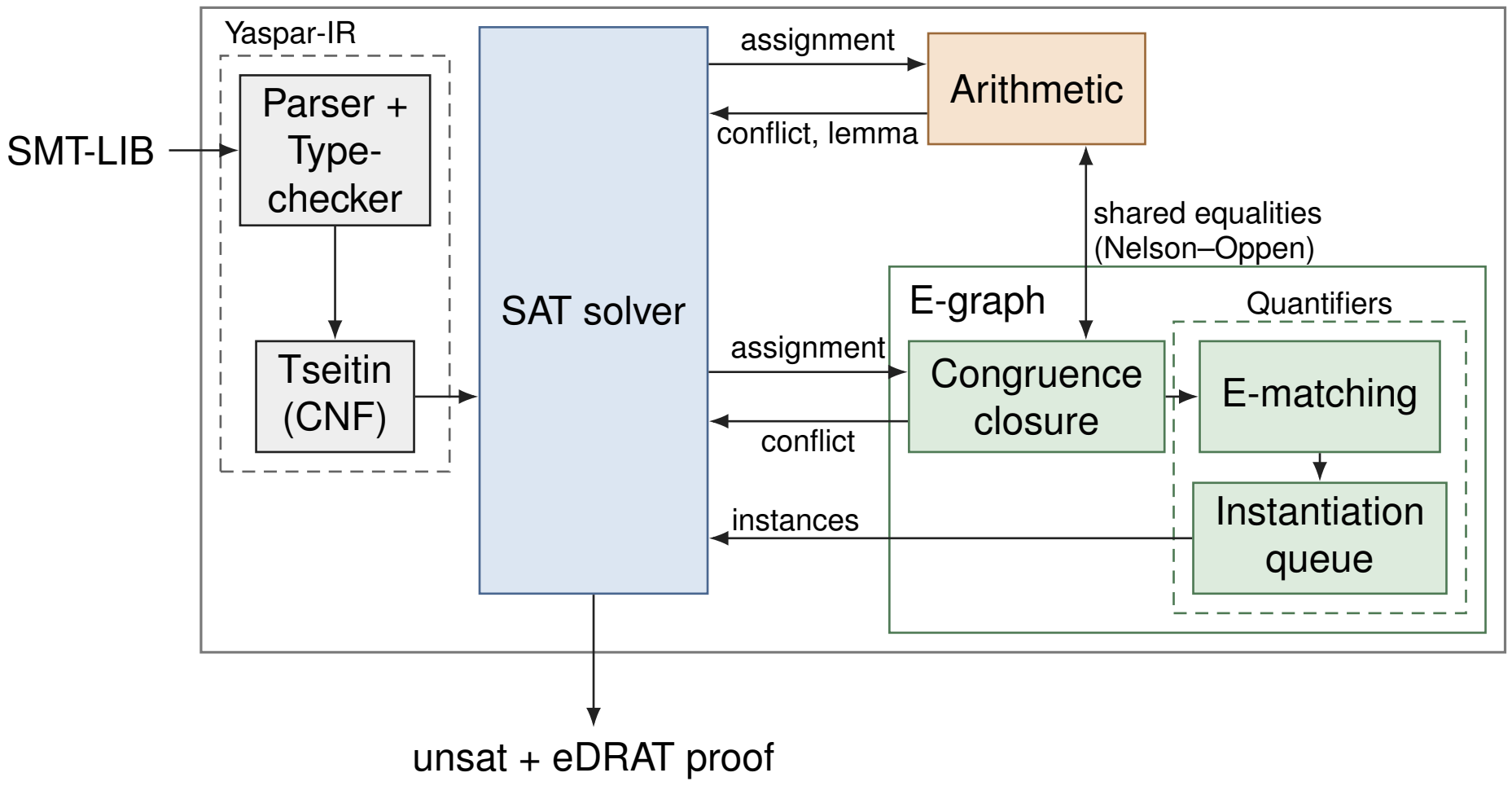
\begin{figure}[t]
\centering
\definecolor{sdsatfill}{HTML}{DCE6F2}\definecolor{sdsatline}{HTML}{4A6790}
\definecolor{sdarfill}{HTML}{F6E3CF}\definecolor{sdarline}{HTML}{9C6A3A}
\definecolor{sdegfill}{HTML}{DDEBDD}\definecolor{sdegline}{HTML}{4F7A55}
\definecolor{sdgray}{HTML}{EEEEEE}\definecolor{sdink}{HTML}{222222}
\begin{tikzpicture}[
    font=\fontfamily{phv}\selectfont\footnotesize,
    >={Latex[length=1.6mm, width=1.2mm]},
    box/.style={draw, line width=0.5pt, minimum height=9mm, inner xsep=5pt,
                align=center},
    lbl/.style={font=\fontfamily{phv}\selectfont\scriptsize, inner sep=1.5pt},
    flow/.style={->, line width=0.5pt, draw=sdink},
  ]

  \node[box, fill=sdsatfill, draw=sdsatline, minimum width=18mm,
        minimum height=46mm]                    (sat)   at (0, 1.25)   {SAT solver};

  \node[box, fill=sdegfill, draw=sdegline]      (cc)     at (3.6, 0.55) {Congruence\\closure};
  \node[box, fill=sdegfill, draw=sdegline]      (ematch) at (5.9, 0.55) {E-matching};
  \node[box, fill=sdegfill, draw=sdegline]      (queue)  at (5.9, -0.6) {Instantiation\\queue};

  \begin{scope}[on background layer]
    \node[draw=sdegline, densely dashed, line width=0.5pt, inner sep=1.5mm,
          fit=(ematch)(queue)] (qi) {};
  \end{scope}
  \node[anchor=south, inner sep=1pt, font=\fontfamily{phv}\selectfont\scriptsize]
       (qilbl) at (qi.north) {Quantifiers};
  \node[anchor=base west, inner sep=0pt] (eglbl) at (cc.west |- qilbl.base) {E-graph};

  \begin{scope}[on background layer]
    \node[draw=sdegline, line width=0.5pt, inner sep=1.5mm,
          fit=(cc)(qi)(qilbl)(eglbl)] (eg) {};
  \end{scope}

  \node[box, fill=sdarfill, draw=sdarline]      (arith)  at (3.6, 3.05) {Arithmetic};

  \node[box, fill=sdgray, draw=sdink]           (cnf)    at (-2.1, 0.55) {Tseitin\\(CNF)};
  \node[box, fill=sdgray, draw=sdink]           (parse)  at (-2.1, 2.55) {Parser +\\Type-\\checker};
  \begin{scope}[on background layer]
    \node[draw=sdink!70, densely dashed, line width=0.5pt, inner sep=1.5mm,
          fit=(parse)(cnf)] (yaspar) {};
  \end{scope}
  \node[anchor=south west, inner sep=1pt, font=\fontfamily{phv}\selectfont\scriptsize]
       at (yaspar.north west) {Yaspar-IR};

  \draw[flow] (parse) -- (cnf);
  \draw[flow] (cnf) -- (cnf -| sat.west);

  \draw[flow] ([yshift=2mm]sat.east |- arith.west) -- node[lbl, above] {assignment}
              ([yshift=2mm]arith.west);
  \draw[flow] ([yshift=-2mm]arith.west) -- node[lbl, below] {conflict, lemma}
              ([yshift=-2mm]sat.east |- arith.west);

  \draw[flow] ([yshift=2mm]sat.east |- cc.west) -- node[lbl, above] {assignment}
              ([yshift=2mm]cc.west);
  \draw[flow] ([yshift=-2mm]cc.west) -- node[lbl, below] {conflict}
              ([yshift=-2mm]sat.east |- cc.west);

  \draw[flow] (cc) -- (ematch);
  \draw[flow] (ematch) -- (queue);
  \draw[flow] (queue.west) -- node[lbl, above, pos=0.82] {instances}
              (sat.east |- queue.west);

  \draw[<->, line width=0.5pt, draw=sdink] ([xshift=4mm]arith.south) -- ([xshift=4mm]arith.south |- cc.north)
       node[lbl, right, pos=0.45, align=left] {shared equalities\\(Nelson--Oppen)};

  \begin{scope}[on background layer]
    \node[draw=sdink!60, line width=0.6pt, inner sep=1.5mm,
          fit=(cnf)(yaspar)(sat)(arith)(eg)] (sd) {};
  \end{scope}

  \node[anchor=east] (in) at ([xshift=-2.5mm]sd.west |- parse) {SMT-LIB};
  \node[anchor=north] (unsat) at ([yshift=-6mm]sat.south |- sd.south) {unsat + eDRAT proof};
  \draw[flow] (in) -- (parse);
  \draw[flow] (sat.south) -- (unsat.north);
\end{tikzpicture}
\vspace{-10px}
\caption{The architecture of \toolname. %
}
\label{fig:sundance-design}
\end{figure}

\toolname is based on the CDCL(T) approach~\cite{smtbookchapter}.  Its
architecture is shown in \autoref{fig:sundance-design}. The input
script is parsed and type-checked by the Yaspar-IR
library~\cite{yaspar,yaspar-ir} and then converted to conjunctive
normal form (CNF). A SAT solver searches for an assignment
to the Boolean skeleton, and theory modules check that assignment
against their theories. A module that finds the assignment
inconsistent returns a \emph{conflict clause}; a module may also
return \emph{lemmas}, clauses that are valid in its theory and guide
the search. The SAT solver adds these clauses and continues, until
either all modules accept a full assignment or the SAT solver derives
the empty clause and returns \texttt{unsat}.

\toolname has three modules. The \emph{e-graph} stores every term in
the query up to equality and implements congruence closure for
uninterpreted functions. The e-graph is also the substrate for
\emph{quantifier instantiation}: e-matching searches it for terms that
match quantifier patterns, and the resulting instances are queued and
handed to the SAT solver. The \emph{arithmetic} module decides linear
integer and real arithmetic using simplex and branch-and-bound.

Inspired by our goals of transparency, trustworthiness, and modularity, we followed
several high-level design principles:

\begin{description}
  \item[Minimalism.] \toolname supports only the SMT fragment
  that program verifiers use. It omits model-based and enumerative quantifier
  instantiation (which verifiers typically disable) and performs little
  preprocessing. We add standard SMT heuristics and optimizations only when they
  give a significant benefit on verification queries.
  \item[Explicit interfaces.] Each module implements a Rust trait and
    can be replaced by any component implementing the same trait. For
    example, \toolname supports using the MiniSat
    solver~\cite{minisat}, Z3's arithmetic
    solver~\cite{bjorner2024arithmetic}, and the Semper e-graph~\cite{semper}
    as alternatives.
  \item[Proof generation.] Every theory lemma and quantifier instance
    reaches the SAT solver as a clause, and \cadical logs these
    together with its own learned clauses. Proof logging thus does not change the search and adds minimal
    overhead.
\end{description}

\subsection{Modules}
\label{subsec:modules}

We now describe each module's role in proving the running example, the
challenge it faces on verification queries, and the approach \toolname takes,
which later sections expand on.

\subsubsection{Parsing.}
\label{subsec:parsing}
\toolname reads its input with Yaspar~\cite{yaspar}, a parser for SMT-LIB~2.7
and represents formulas with Yaspar-IR~\cite{yaspar-ir}. Yaspar-IR
type-checks the parsed script and produces a typed, hash-consed
representation of terms. Yaspar and Yaspar-IR are standalone libraries
that other tools can reuse.

\subsubsection{SAT Solver.}
\label{subsec:sat}
Using Yaspar-IR, \toolname converts the query to CNF using the Tseitin
transformation~\cite{tseitin}, and the SAT solver treats each \emph{atom}
as a separate SAT variable. In our example, the SAT solver makes the
case split of the proof, on whether $j_0 < i$ and on
whether $\mathit{index}(a,i) > \mathit{ub}$, which selects the branch
of the \smtcode{ite} that defines $\mathit{ub\_new}$.

As the SAT solver picks a truth assignment for the atoms, it notifies
the theory modules of each assignment. Theory modules can check each
partial assignment \emph{eagerly}, as literals are assigned, or
\emph{lazily}, i.e., once the SAT solver has a complete assignment. In
\toolname, the e-graph and arithmetic modules are eager and quantifier
instantiation is lazy.

In existing SMT solvers, the SAT solver is typically %
tightly coupled to the theory solvers, and its many heuristics are tuned
together with theirs. This makes it hard to adopt advances in SAT solving, or
to tell which heuristics actually matter for verification queries.
\toolname instead uses an off-the-shelf SAT solver through the IPASIR-UP
interface~\cite{ipasir-up}, and currently supports both
\cadical~\cite{cadical} and MiniSat~\cite{minisat}. 
Modules never directly influence SAT search; they only add clauses.
This keeps the
interface small, so we can experiment with SAT-level techniques such as
chronological backtracking and phase selection independently of the theory
modules (\autoref{sec:satsolver}).

\subsubsection{E-graph.}
\label{subsec:egraph}

In our example, the e-graph carries out the last step of the second
case: once $j_0 = i$ is known, congruence concludes $\mathit{index}(a,
j_0) = \mathit{index}(a, i)$. The e-graph implements congruence
closure~\cite{congruence-closure,nelson-thesis}, and it is also the
shared data structure through which Booleans, \smtcode{ite} terms,
datatypes, and quantifier matching are handled.
There are many e-graph designs, from the backtrackable congruence closure of
SMT solvers to e-graph libraries built for equality
saturation~\cite{egg,egglog,semper}, and they make different trade-offs between
efficiency, simplicity, and expressivity. In existing SMT solvers, the e-graph
is deeply integrated with the rest of the solver, so these trade-offs are hard
to explore.

\toolname places the e-graph behind a trait, so alternative implementations can
be substituted, and we used this to experiment with several design choices. We
keep every term in the e-graph, including Boolean terms; index congruences with
a signature table; and keep constants as representatives of their equivalence classes so that
conflicting merges are detected cheaply (\autoref{sec:egraph}).

\subsubsection{Quantifier Instantiation.}
\label{subsec:quantifiers}
In our example, quantifier reasoning carries out the first two
steps of the proof: naming $j_0$ skolemizes the negated goal, introducing a fresh
constant, and applying the invariant to $j_0$ instantiates it with $j := j_0$.
Since the quantifier has infinitely many instances, a solver must decide both
\emph{how} to discover candidate instances and \emph{when} to add them to the
search. Existing solvers answer both questions with many interacting heuristics,
and on large verification queries these heuristics largely determine
performance.
To discover instances, \toolname uses top-down e-matching~\cite{simplify} on the
patterns supplied by the verifier: the pattern \smtcode{(index a j)} matches
$\mathit{index}(a, j_0)$ in the e-graph, yielding the needed instance. We found
that the simple approach is sufficient for our Verus benchmarks, as the time
spent searching for quantifier instantiations is not a significant bottleneck
there.
To decide when to instantiate, \toolname matches in rounds
against a fixed e-graph and adds the resulting instances to the SAT solver
lazily, in batches.

\subsubsection{Arithmetic.}
\label{subsec:arith}
Both cases of the proof end in arithmetic conflicts, and the second relies on
integer reasoning: $i \le j_0 < i + 1$ forces $j_0 = i$. The challenge is theory
combination: the arithmetic module and the e-graph must agree on which terms are
equal, but sharing every candidate equality floods the SAT solver with lemmas.
\toolname shares equalities lazily: it tentatively merges terms that are
equal in the arithmetic model, and only adds a lemma when the merge causes a
conflict in the e-graph. The arithmetic backend is itself a module: \toolname
can use its own built-in arithmetic module or use Z3 for arithmetic (\autoref{sec:arithmetic}).

\subsubsection{Proof Production.}
\label{subsec:proofs-overview}
When \toolname reports that the example is \texttt{unsat}, it also produces a
proof that can be checked independently.
In existing SMT solvers, proof production places heavy demands on
every component: each theory solver and preprocessing step must justify its
reasoning in a fine-grained proof
calculus~\cite{cvc5-proof,alethe}. This is costly to implement, slows
solving, and can leave gaps where a component does not support
proofs.
\toolname instead uses eDRAT~\cite{edrat}, a clausal proof format in the style
of SAT solvers. Because every module communicates with the SAT solver only
through clauses, the proof is just a record of these clauses: \cadical logs its
learned clauses, and each theory lemma or quantifier instance is logged as it is
added. Such proofs are easy to produce and easy to check: the propositional
steps are checked by a SAT proof checker and each theory lemma is checked separately. They
also make it easy to swap in new components, since a module only needs to
produce valid clauses (\autoref{sec:proofs}).

\paragraph{Measuring design decisions.}
In \autoref{sec:egraph} and \autoref{sec:quantifiers}, we report how individual
design decisions affected \toolname's performance during development. We
measured each change by rerunning the published version of our Verus
development benchmark (\autoref{sec:eval}) with a 60\,s timeout
before and after the change. These numbers reflect
\toolname at the time of each change, not its final version.

\section{SAT Solver}\label{sec:satsolver}
\toolname delegates Boolean search to an off-the-shelf SAT solver.
\toolname communicates with the SAT solver through the IPASIR-UP
interface~\cite{ipasir-up}, using callbacks that allow its e-graph and
arithmetic modules to eagerly process assignments to Boolean
variables, decision levels, and backtracks.  When the SAT solver finds
a complete satisfying assignment to the Boolean skeleton, \toolname
performs datatype reasoning, further arithmetic checks, and runs a
quantifier round. If the Boolean model is theory-inconsistent or a
quantifier round produces new instances, \toolname adds clauses to
the SAT solver: conflict clauses, theory lemmas, and clauses for Skolem
constants and quantifier instances.

The default SAT solver is \cadical, but as an alternative, \toolname
also supports MiniSatUP~\cite{minisatup-github}, a version of
MiniSat that exposes an IPASIR-UP interface. By design, the SAT solver
can be replaced by any IPASIR-UP solver without requiring changes to
theory modules.  However, performance depends on specific
SAT-solver features.
\toolname adds many clauses, especially from quantifier instantiation, so the
\texttt{elevate} option added in \cadical 3.0.1~\cite{cadical301} is critical to
avoid discarding large parts of the current assignment upon each addition
(\autoref{subsec:q-schedule}); similar functionality is built into the version
of MiniSat that we use. Additionally, setting the initial phase for new
variables to false results in degraded performance on many benchmarks, so we configure our version
of MiniSat to use true as the initial phase, matching the default phase
setting in \cadical.  With this setting, MiniSat solves only slightly fewer
Verus queries than \cadical (\refappendix{app:sat}).

\section{E-graphs}\label{sec:egraph}

The e-graph is \toolname's central data structure. It decides equality
with uninterpreted functions through congruence closure. It is
also the term index that e-matching searches and the channel through
which arithmetic and datatype reasoning learn about equalities. We
first describe how the e-graph computes and explains congruences, then
the design decisions that shaped it, and finally how it interacts with
the rest of \toolname.

\subsection{Congruence Closure and Explanations}\label{subsec:eg-cc}
\label{subsec:eg-explain}

\toolname's congruence closure follows the design of Nieuwenhuis and
Oliveras~\cite{nieuwenhuis2005proof}. It extends a union-find data
structure with a proof forest: each node has at most one outgoing
edge, labeled with the reason its endpoints are equal, and the root of
each tree represents an e-class. An \emph{equality} edge records an
equality asserted by the SAT solver; a \emph{congruence} edge records
the argument pairs that made two applications congruent. To merge the
classes of $x$ and $y$, the e-graph re-roots $y$'s tree at $y$ and
adds an edge from $y$ to $x$, so that edges always connect terms that
were actually equated. It then checks the disequalities stored at the
surviving root and looks for new congruences among the parents of
$y$'s old root.

\begin{figure}[t]
\centering
\begin{subfigure}[b]{0.34\textwidth}
\centering
\small
$\begin{array}{ll}
(1) & e = u \\
(2) & e = g(f(a)) \\
(3) & a = b \\
(4) & g(f(b)) \neq u
\end{array}$
\vspace{6mm}
\subcaption{Asserted literals.}\label{fig:cc-example-lits}
\end{subfigure}%
\hfill
\begin{subfigure}[b]{0.62\textwidth}
\centering
\definecolor{lcaS}{HTML}{2F5D9E}  %
\definecolor{lcaT}{HTML}{B5651D}  %
\begin{tikzpicture}[
    font=\small,
    >={Latex[length=1.6mm, width=1.2mm]},
    t/.style={draw, line width=0.5pt, rounded corners=1pt, inner sep=3pt,
              minimum height=6mm, fill=white},
    eqe/.style={->, line width=0.9pt},
    cge/.style={->, line width=0.9pt, dash pattern=on 3pt off 2pt},
    lbl/.style={font=\footnotesize, inner sep=2pt},
  ]

  \node[t, line width=1.4pt]  (e)  at (1.1, 2.2) {$e$};
  \node[t]                    (g1) at (0.0, 1.1) {$g(f(a))$};
  \node[t]                    (u)  at (2.2, 1.1) {$u$};
  \node[t]                    (g2) at (0.0, -0.2) {$g(f(b))$};

  \draw[eqe, lcaS] (g1) -- node[lbl, left]  {(2)} (e);
  \draw[eqe, lcaT] (u)  -- node[lbl, right] {(1)} (e);
  \draw[cge, lcaS] (g2) -- node[lbl, left]  {$f(b){=}f(a)$} (g1);

  \node[font=\footnotesize, anchor=west] at ([xshift=1mm]e.east) {LCA};

  \node[t] (fa) at (3.6, 1.1) {$f(a)$};
  \node[t] (fb) at (3.6, -0.2) {$f(b)$};
  \node[t] (a)  at (4.7, 1.1) {$a$};
  \node[t] (b)  at (4.7, -0.2) {$b$};

  \draw[cge, lcaS] (fb) -- node[lbl, left]  {$b{=}a$} (fa);
  \draw[eqe, lcaS] (b)  -- node[lbl, right] {(3)} (a);
\end{tikzpicture}
\subcaption{Proof forest after (1)--(3).}\label{fig:cc-example-forest}
\end{subfigure}
\caption{Explaining a conflict. Arrows point from a node to its parent. Solid
edges record an asserted equality; dashed edges are congruence edges, labeled
with the argument pair they depend on. To explain $g(f(b)) = u$, the e-graph
follows the blue path from $g(f(b))$ and the orange path from $u$ to their
least common ancestor $e$, and recursively explains the congruence edges
(also blue).}
\label{fig:cc-example}
\end{figure}
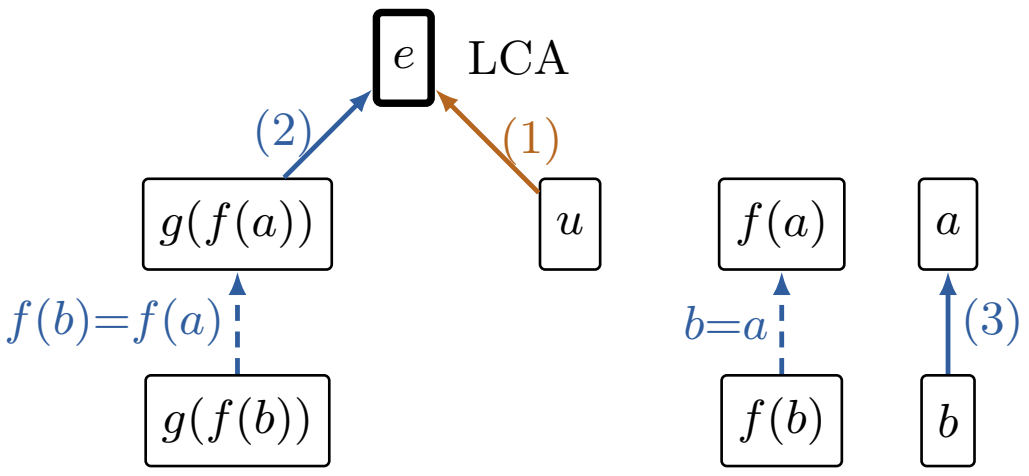

When a merge violates a disequality $s \neq t$, the e-graph must explain why
$s = t$ in terms of literals the SAT solver assigned in order to learn a
conflict clause. In
\autoref{fig:cc-example}, literals (1) and (2) put $u$ and $g(f(a))$ in the
class of $e$; (3) makes $f(a)$ and $f(b)$ congruent, and hence $g(f(a))$ and
$g(f(b))$, which violates (4). Since $g(f(b))$ and $u$ are in the same tree,
they are connected by a single path.
Each edge on this path contributes to the explanation: an equality edge
contributes its literal, (2) and (1) here, and a congruence edge
contributes the explanations of its argument pairs, computed
recursively in the same way. Explaining $f(b) = f(a)$ yields (3), so
the conflict clause is
\[
  \lnot(e = u) \lor \lnot\bigl(e = g(f(a))\bigr) \lor \lnot(a = b) \lor
  g(f(b)) = u.
\]
Although the conflict involves two congruence steps, the clause mentions only
literals the SAT solver assigned. When an argument pair has no equality atom in
the input, \toolname introduces one. These clauses enter the eDRAT proof as
theory lemmas (\autoref{sec:proofs}).

\subsection{Design Decisions}\label{subsec:eg-decisions}

Several of these choices are standard in congruence closure; we include them
because they had a large effect on performance or simplicity.

\paragraph{All terms in the e-graph.}\label{subsec:eg-terms}
Every term is an e-graph
node, including Boolean terms. When the SAT solver assigns an atom, the e-graph
merges it with \smtcode{true} or \smtcode{false}. This lets congruence reason
about predicates and \smtcode{ite} terms directly: for example, if $a = b$,
$P(a)$ is true, and $P(b)$ is false, congruence merges \smtcode{true} with
\smtcode{false}, which is a conflict. It also allows pattern matching against
Boolean terms. Keeping
Boolean terms out of the e-graph could be more efficient, but in our experiments
excluding them gave only a negligible speedup.

\paragraph{Signature table.}
To find new congruences, the e-graph keeps a \emph{signature table} mapping
each signature $(f, \mathit{find}(t_1), \ldots, \mathit{find}(t_n))$, where
$\mathit{find}(t)$ is the representative of $t$'s class, to an application with
that signature. After a merge, each application with an argument in the merged
class is looked up in the table: a hit in a different class is a new
congruence, merged immediately. Adding the signature table increased the number
of Verus queries \toolname solved by 3.7\% on our development benchmark
(\autoref{subsec:modules}).

\paragraph{Constants as representatives.}
A constant, such as \smtcode{3} or \smtcode{true}, is always kept as
the representative of its class. A merge of two classes with different
constant representatives, such as \smtcode{true} and \smtcode{false},
is therefore detected as a conflict with a single comparison. This change increased the number of Verus queries
\toolname solved by 9.9\%, the largest single
improvement during development.

\paragraph{No theory propagation.}
\toolname reports conflicts but never propagates implied literals to the SAT
solver. We experimented with theory propagation and found that it made no
significant difference on program verification benchmarks, so we left it out
for simplicity.

\paragraph{A modular e-graph.}
The rest of the solver accesses the e-graph only through a Rust trait covering
term registration, asserting and querying (dis)equalities, explanations,
pattern matching, and backtracking, so a different implementation, such as the
Semper e-graph~\cite{semper}, can be easily substituted.

\paragraph{Integration with Semper.}
Semper~\cite{semper} is an e-graph providing the same core operations as
\toolname's e-graph, but designed for equality saturation
workloads~\cite{eqsat}. Specifically, Semper is optimized for marking and restoring
very large e-graphs. To save space, it compresses the changes made since the
last mark incrementally, and decompresses them when a restore
undoes them. 

Semper also recycles the identifiers of nodes that a restore has released. A
restore deletes every node created since the mark, and later nodes may reuse
their identifiers. After each restore, the adapter therefore
re-inserts every term registered above the restore point and updates the term
table to the new nodes. This re-insertion takes 56\% of a quantifier instantiation
round, and it accounts for much of the slowdown of \toolname with Semper
(\autoref{subsec:eval-modules}).

\subsection{Interaction with Other Modules}\label{subsec:eg-interface}
\label{subsec:eg-backtrack}

\paragraph{Backtracking.}
The e-graph processes each literal as soon as the SAT solver assigns it. It must
undo its state whenever the SAT solver backtracks. Every merge is recorded on a trail and reversed on backtracking. Stale parent lists
and disequality entries are invalidated lazily rather than deleted. Terms
introduced during search, by quantifier instances or datatype axioms, are
permanent; after backtracking, \toolname re-inserts them so their
congruences are rediscovered.

\paragraph{Arithmetic.}
Every merge between classes containing arithmetic terms is sent to the arithmetic
solver as an equality. In the other direction, \toolname uses a model-based
variant of Nelson--Oppen combination~\cite{smtbookchapter,yices2}: when the
arithmetic solver accepts a full assignment, \toolname tentatively merges pairs
of e-classes with the same value in the arithmetic model. Only if a merge causes
a conflict does it add the lemma $x < y \lor x = y \lor x > y$; all tentative
merges are then undone. This generates lemmas only for equalities that actually
matter to the e-graph.

\paragraph{Datatypes.}
Like Z3~\cite{z3-internals}, \toolname adds some datatype axioms
eagerly when a datatype term is registered: for a constructor application
$C(t_1, \ldots, t_n)$, it asserts $\mathit{is}\text{-}C(C(t_1, \ldots, t_n))$
and $\mathit{sel}_i(C(t_1, \ldots, t_n)) = t_i$, and for terms of non-recursive
datatypes it asserts that some tester holds. When a tester
$\mathit{is}\text{-}C(t)$ is assigned true, \toolname lazily adds the axiom
$t = C(\mathit{sel}_1(t), \ldots, \mathit{sel}_n(t))$. Acyclicity is checked at
final check by a depth-first search over e-class representatives; a cycle
yields a conflict explained by the e-graph equalities along it.

\section{Quantifier Instantiation}\label{sec:quantifiers}

Program verification queries are dominated by quantifiers. Verifiers
encode function definitions, datatype and collection axioms, and loop
invariants as universally quantified formulas with syntactic patterns, and the
solver's job is mostly to find the instances that close the
proof. \toolname supports only the style of quantifier reasoning that
verifiers rely on: pattern-based instantiation by
e-matching~\cite{simplify,e-matching}, and skolemization of
existential quantifiers. After describing how quantifiers are
represented, we discuss the two questions raised in
\autoref{subsec:quantifiers}: how \toolname discovers instances, and
when it adds them to the search.

\subsection{Quantifiers as Literals}\label{subsec:q-literals}
\label{subsec:q-skolem}

\toolname does not preprocess quantifiers. Each quantified formula is a Boolean
atom like any other: the SAT solver assigns it, and the e-graph stores it as an
opaque term. A universal $P = \forall \vec{x}.\, \varphi(\vec{x})$ assigned
true, or an existential $Q = \exists \vec{x}.\, \varphi(\vec{x})$ assigned
false, is \emph{instantiated}. A universal  assigned false, or an existential
assigned true, is \emph{skolemized}.

\subsection{Discovering Instances}\label{subsec:q-ematch}

\paragraph{Patterns.}
Each universal quantifier carries one or more \emph{multi-patterns}: lists of
terms over the bound variables that must all match under a single substitution.
Program verifiers annotate almost all quantifiers with patterns. For inputs
without them, \toolname can optionally infer patterns in the style of
Simplify~\cite{simplify}, preferring a single term that covers all bound
variables and otherwise choosing a small multi-pattern by greedy set cover.

\paragraph{E-matching.}
\toolname implements the top-down backtracking e-matching algorithm of
Simplify~\cite{simplify}. For a pattern $f(p_1, \ldots, p_n)$, it scans the
registered applications of $f$ and matches each argument recursively, modulo
the current e-classes: a pattern variable binds to the argument's e-class, a
ground subterm must be in the argument's class, and a nested application is
matched against the applications of its symbol in that class.

\paragraph{Matching is not the bottleneck for Verus.}
More efficient algorithms exist. Z3 compiles patterns into code
trees~\cite{e-matching} that share work across patterns, and relational
e-matching~\cite{egg-relational} offers asymptotic speedups on complex patterns;
we also experimented with the latter. On our Verus development benchmark,
however, matching takes a small fraction of solving time, so the simple
algorithm sufficed during development. This is not true of every benchmark: on
the very large \fstar queries from DICE$^\star$, matching does become a
bottleneck (\autoref{subsec:eval-mariposa}). Because matching is a separate
module, a faster algorithm can be substituted in the future
(\autoref{sec:futurework:quantifierheuristics}).

\subsection{Scheduling Instances}\label{subsec:q-schedule}

\paragraph{Rounds against a fixed e-graph.}
\toolname runs quantifier reasoning lazily, only when the SAT solver finds a full
assignment that the e-graph and arithmetic module accept. It then runs a
\emph{round}: it skolemizes every newly assigned quantifier that requires it and
matches the patterns of every universal quantifier assigned true. All matches
are computed before any instance is added to the e-graph, so a round's matches
depend only on the e-graph at its start, not on the order in which instances
are added. We found this to be faster and more robust: it increased the number
of Verus queries \toolname solved by 0.6\% on our
development benchmark (\autoref{subsec:modules}).

\paragraph{Lazy batches.}
Matches are placed in a pending queue, skipping any instance already produced.
\toolname adds up to a fixed batch of pending instances (85 by default) to
the SAT solver, and the rest one per subsequent full assignment, starting a new round
only once the queue is empty. Bounding the batch lets the SAT solver react to
the first instances, which often produce a conflict, before many more clauses
are added. Adding a clause rejects the current assignment, so \toolname uses
\cadical's \texttt{elevate} option, introduced in
\cadical~3.0.1~\cite{cadical301}, to backjump only as far as needed and keep
most of the assignment between batches. Together, these changes increased the
number of Verus queries \toolname solved by 2.1\% on the
same benchmark.

\section{Arithmetic}
\label{sec:arithmetic}

\toolname includes its own solver for mixed linear integer and real
arithmetic, based on techniques by Dutertre and de
Moura~\cite{dutertre2006fast} and Bromberger~\cite{bromberger2019decision}. It
can be used as a theory solver or as a standalone \texttt{QF\_LIRA} solver.
Like the rest of \toolname, it favors simplicity and stability over peak
performance, targeting the small to medium linear systems that arise in program
verification. It consists of a preprocessor, a dual simplex solver, and integer
reasoning on top of simplex.

\subsection{Preprocessing}

The solver receives arithmetic literals from the SAT solver and e-graph and applies only
lightweight preprocessing. It orders the variables deterministically, moves
variables to one side and constants to the other, and scales each relation so
that its leading coefficient is $1$; relations with the same linear term can
then share a slack variable~\cite{dutertre2006fast}. Trivially true relations,
such as $0 < 1$, are dropped, and a trivially false relation, such as $0 < -1$,
is immediately reported as a conflict. Equalities become a pair of bounds, and
disequalities are handled by the e-graph, which avoids splitting $x \neq y$
into $x < y \lor x > y$. The result is a set of bounded linear sums
\begin{equation}
l_i \le \sum_{j} a_{ij} x_j \le u_i
\label{eq:preproc}
\end{equation}
with rational coefficients and bounds that are rational or $\pm\infty$.

\subsection{Simplex}

The rational relaxation is decided by a dual simplex solver
that closely follows existing
implementations~\cite{dutertre2006fast,bromberger2019decision}. Strict
inequalities are handled by working over $\mathbb{Q}_\delta$, replacing
$l < x$ with $l + \delta \le x$ for an infinitesimal $\delta$, which is
instantiated with a suitable rational value when a model is produced. Each
relation in \autoref{eq:preproc} receives a slack variable $s_i$, giving the
tableau $\mathbb{A}x = s$ with bounds $l_i \le s_i \le u_i$. The solver then
runs the standard dual simplex algorithm~\cite[Sec.~2.7.1]{bromberger2019decision},
using Bland's rule~\cite{bland1977new} to choose pivots and guarantee
termination. We chose Bland's rule over more sophisticated pivoting heuristics~\cite[Ch.~7]{bromberger2019decision}
for simplicity and determinism. The dual simplex solver is
incremental: bounds can be asserted in any state, and the solver can set and
return to backtrack points.

\subsection{Integer Reasoning}

When the rational relaxation is feasible but some integer variable has a
non-integral value, the solver first tries cheap tests that can decide the
integer problem directly: the GCD test~\cite[Sec.~4.3]{bjorner2024arithmetic}
and the fast cube test~\cite[Ch.~4]{bromberger2019decision}. If these fail, it
falls back on branch-and-bound~\cite[Sec.~2.7.3]{bromberger2019decision}, using
the incremental interface to explore strengthened bounds. Branch-and-bound is
not complete for unbounded problems, so the solver may time out; Bromberger's
bounding transformation~\cite{bromberger2019decision}, which turns any mixed
integer system into an equisatisfiable bounded one, would remove this
limitation and is left for future work. The solver also omits techniques
common in other SMT solvers~\cite{bjorner2024arithmetic}, such as Gomory cuts,
Hermite normal forms, and nonlinear arithmetic.

The end-to-end arithmetic module exposes an incremental interface. This is driven
by the SAT solver and used to detect conflicts early as the SAT solver builds
partial models. The implementation relies on a growing tableau with slack row
caching to avoid tableau size blow-up. When assertions are backtracked over, the
corresponding slack variable's bounds are loosened to $\pm\infty$, preserving
satisfiability while not requiring the row to be removed or any pivots to occur.

\subsection{Modularity}

The arithmetic module implements a Rust trait, so it can be replaced without
changing the rest of \toolname. We have also implemented this trait using Z3's
arithmetic solver through its Rust API; the two backends are selected with a
command-line flag. 

\section{Proof Production}
\label{sec:proofs}

\toolname produces proofs in the eDRAT format~\cite{edrat}, an
extension to SMT of the DRAT format used by SAT solvers.  eDRAT uses
SMT-LIB syntax for declaring sorts, functions, and constants, and for
defining terms and formulas. In addition, an eDRAT proof records
theory and Boolean reasoning steps performed by SMT solvers in the
form of clauses. A clause is written in the DIMACS style as a list of
integers. eDRAT distinguishes three types of clauses: assumptions
correspond to the assertions in the original SMT file after conversion
to CNF, theory lemmas are clauses derived by theory reasoning, and
regular clauses are derived by Boolean resolution. As in DRAT,
the format also allows clauses to be deleted. eDRAT proofs can be
produced by \toolname with negligible overhead.

We have modified the original eDRAT format described in~\cite{edrat}
to allow theory lemmas to be labeled by a theory identifier. For
example, theory lemmas produced by congruence closure and lemmas
produced by the arithmetic solvers have different labels, which
facilitates proof checking. We have also added new constructs to
describe skolemization and quantifier instantiation.

Proof checking is done by using the Valido toolchain described
in~\cite{edrat}. We have extended Valido to check quantifier reasoning
steps and to parse the modified eDRAT format. We have also generalized
Valido to use external SMT solvers to validate theory
lemmas.\footnote{We are in the process of open-sourcing Valido.}

\section{Evaluation}
\label{sec:eval}

In this section, we aim to answer three research questions:

\begin{enumerate}
    \item[\textbf{RQ1}] How does \toolname compare with existing SMT solvers on
    Verus, which it was initially developed for?
    \item[\textbf{RQ2}] How well does \toolname generalize to other verification
    frameworks like Dafny and \fstar?
    \item[\textbf{RQ3}] How does \toolname's performance change when swapping out its modules?
\end{enumerate}

\paragraph{Benchmarks.}

We evaluate on several datasets. First, we evaluate on a set of 2,576 Verus
queries, originally compiled by Zhou et al.~\cite{cazamariposas}, from a number
of Verus systems verification projects: Anvil~\cite{anvil},
Atmosphere~\cite{atmosphere}, Verified SplinterDB~\cite{splinterdb},
VeriSMo~\cite{verismo}, and a collection of other Verus systems verification
projects~\cite{verus-sys}. A subset of this benchmark set was first submitted to
the SMT competition in 2025~\cite{smtcomp2025verus}. We used this benchmark set
to guide the development of \toolname, and the effects of design decisions
reported in \autoref{sec:egraph} and \autoref{sec:quantifiers} were measured on
it. Its published version contains 3,763 queries, of which 1,187 are duplicate
copies of SplinterDB queries; we remove these duplicates here, leaving 2,576. We
directly shared the benchmarks with the developers of cvc5 and Z3 when we
started developing \toolname so that both solvers could improve on Verus benchmarks.

To answer RQ1 without the bias of evaluating on the benchmarks we developed
against, we also created a new, \emph{held-out} set of 4,086 Verus queries that
was not used during development. These come from four more recent Verus
projects: CortenMM~\cite{cortenmm}, OwlC~\cite{owlc}, Verdict~\cite{verdict},
and vostd~\cite{vostd}. To answer RQ2, we use the Dafny and
\fstar queries from the \mariposa benchmark suite~\cite{mariposa}: 11,674 Dafny
queries from Komodo~\cite{komodo} and from two versions of VeriBetrKV, one
using dynamic frames~\cite{dfvbkv} and one using linear
types~\cite{linear-dafny}, and 3,195 \fstar queries from DICE$^\star$~\cite{dice-star}
and vWasm~\cite{wasm-sandboxing}. \toolname was not developed on these.

For the development Verus benchmarks, we pick the UFDTLIA queries.
For the other benchmark sets, we compiled the queries from the original
verification tool and filtered out any queries that used theories that we did
not support such as bit-vectors, nonlinear arithmetic, and strings. Overall, we 
found that we could support 91.3\% of the Verus held-out benchmarks, 97.1\% of
the \fstar benchmarks, and 93.0\% of the Dafny benchmarks.

\paragraph{Setup.}
All experiments run on an AWS EC2 m7i.48xlarge instance (2$\times$ Intel Xeon
Platinum 8488C, 192 hardware threads, 743\,GB RAM), with 150 queries solved in
parallel. Each query has a 60\,s wall-clock timeout, and counts as solved only
if the solver returns \texttt{unsat}. \toolname 
always produces eDRAT proofs, and on Dafny and \fstar queries, which contain
quantifiers without patterns, it infers patterns automatically. We checked the
proof of every \texttt{unsat} answer \toolname produced in our evaluation with
Valido (\autoref{sec:proofs}); every check succeeded except one, which timed
out. We compare
against Z3~5.3.0~\cite{z3}, cvc5~1.4.2 (git \texttt{282d5e0})~\cite{cvc5}, and
Alt-Ergo~2.4.3~\cite{altergo}. Z3 runs with the options each verifier passes to
it. cvc5 runs with \texttt{--no-cbqi --user-pat=strict}, the configuration recommended
by its developers. Both cvc5 and Z3 regress with their default configurations.

Because \toolname always produces proofs, we run Z3 and cvc5 with proof
production enabled. We also report Z3 without proofs, as proof production is a
significant overhead. Alt-Ergo does not produce proofs; it runs with its default
options, but we have to translate the inputs to its SMT-LIB dialect.

We also ran SMTInterpol~2.5~\cite{smtinterpol}, which
supports the logics and features used by Verus, on the two Verus benchmark sets.
It solves only 17.7\% of the development set and 14.2\% of the held-out set
within 60\,s, so we omit it from the rest of the evaluation.

\subsection{RQ1: Verus}\label{subsec:eval-verus}

\begin{figure}[t]
\centering
\begin{subfigure}[t]{0.49\textwidth}
\includegraphics[width=\textwidth]{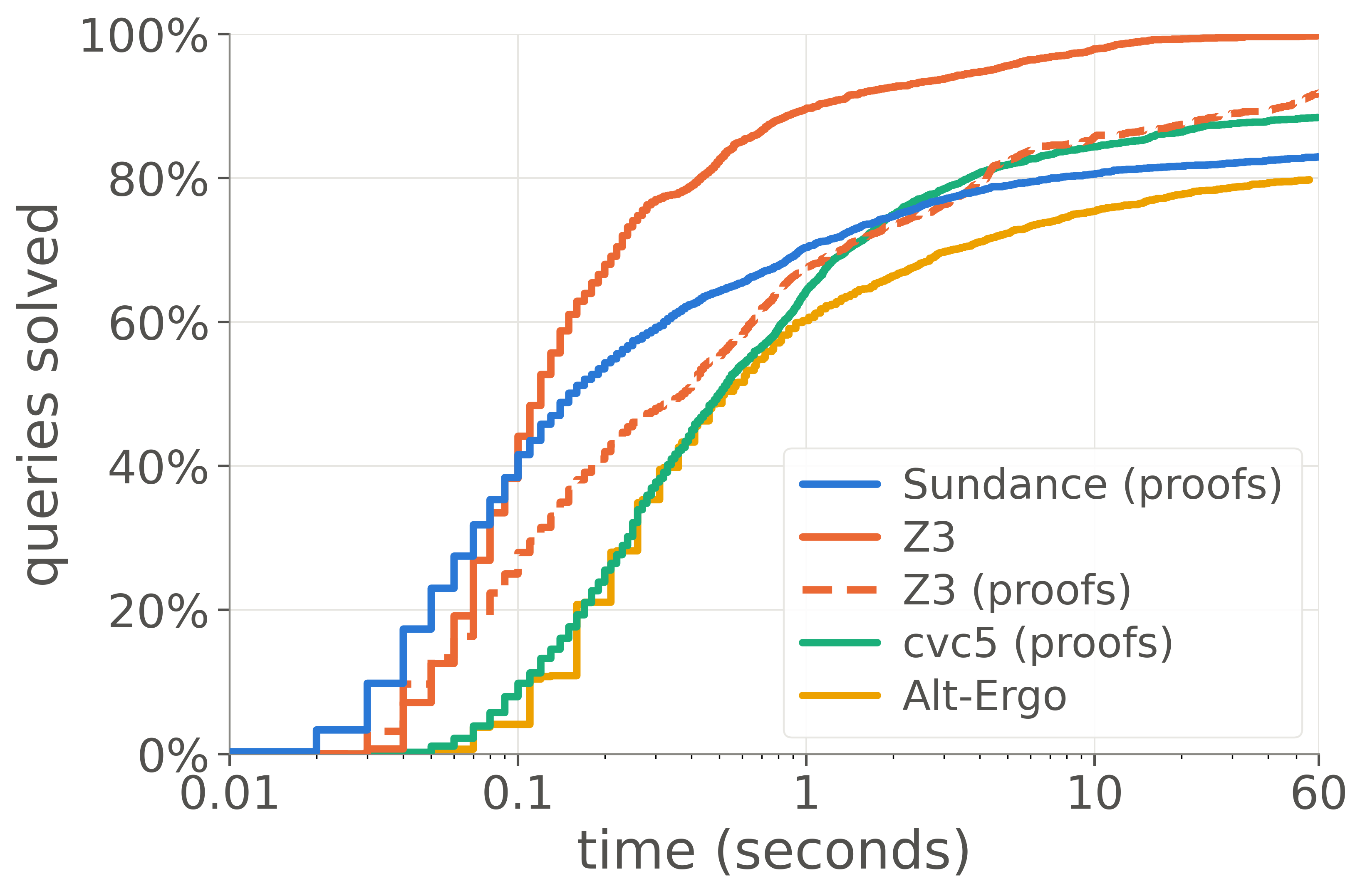}
\subcaption{Development set.}\label{fig:eval-verus-dev}
\end{subfigure}%
\hfill
\begin{subfigure}[t]{0.49\textwidth}
\includegraphics[width=\textwidth]{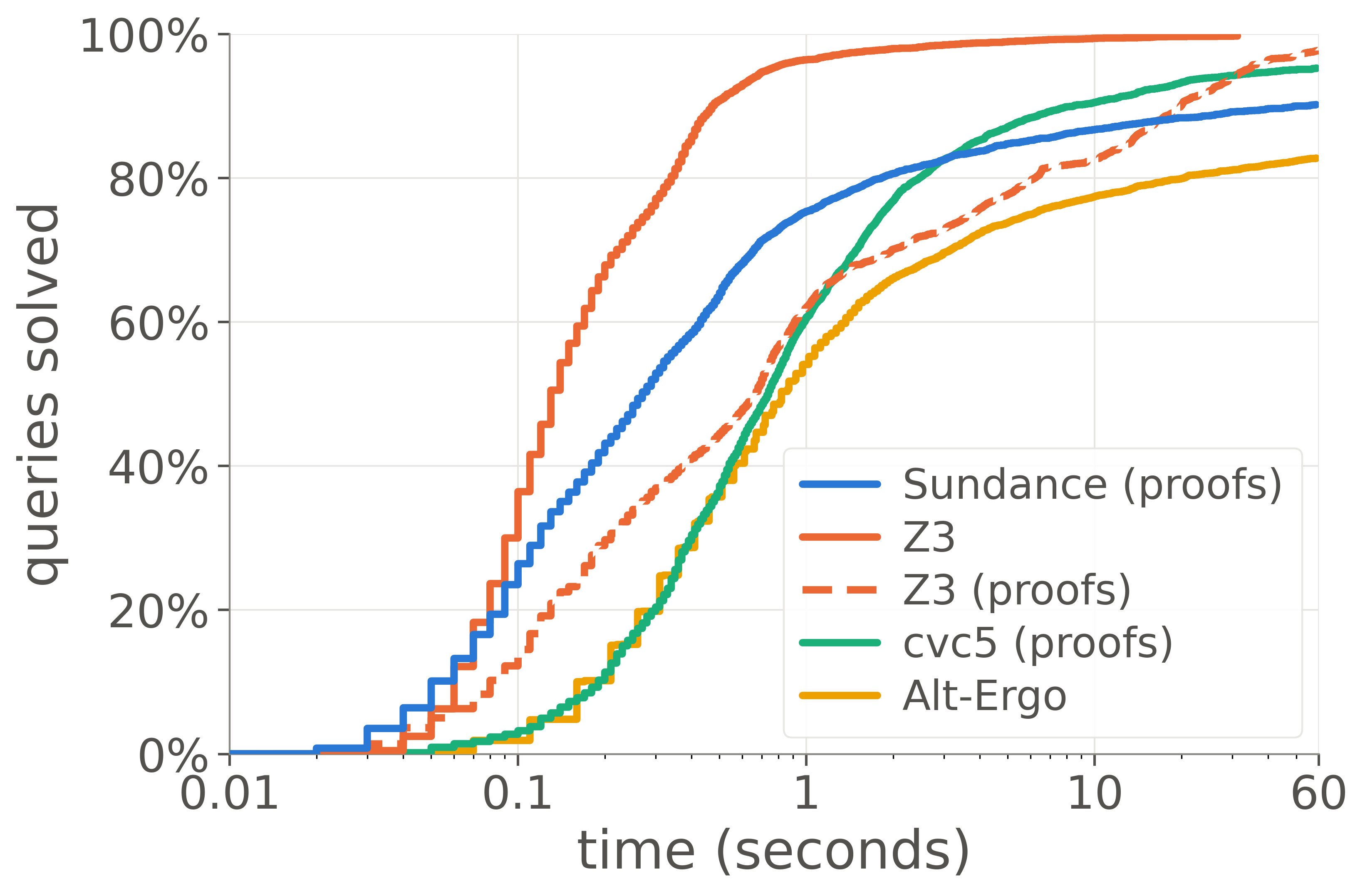}
\subcaption{Held-out set.}\label{fig:eval-verus-heldout}
\end{subfigure}
\caption{Percentage of Verus queries solved within a given time budget.
\toolname, Z3 (proofs), and cvc5 (proofs) produce proofs; Z3 and Alt-Ergo do
not.}
\label{fig:eval-verus}
\end{figure}

\autoref{fig:eval-verus} shows the results on the two Verus benchmark sets.
Apart from Z3, the solvers perform comparably. Within 60\,s, \toolname solves
83.0\% of the development set, compared with 88.4\% for cvc5, 91.8\% for Z3 with
proofs, and 79.8\% for Alt-Ergo; on the held-out set, the corresponding numbers
are 90.2\%, 95.3\%, 97.7\%, and 82.8\%. Z3 without proofs solves 99.8\% of both
sets.

At short time budgets, \toolname is faster than every solver except Z3
without proofs. Within 1\,s, it solves 70.3\% of the development set and
75.4\% of the held-out set, compared with 67.5\% and 62.1\% for Z3 with proofs,
64.4\% and 60.7\% for cvc5, and 60.2\% and 54.1\% for Alt-Ergo. We believe this is because \toolname does minimal preprocessing.

We attribute Z3's lead to two factors. First, Z3 has mature quantifier
instantiation heuristics, such as eager instantiation and relevancy propagation,
that \toolname does not yet implement (\autoref{sec:futurework}). Second, the
benchmarks are biased towards Z3 in two ways. The verifiers themselves are
built around Z3: their encodings, pattern choices, and solver options have
been tuned to its heuristics
(\autoref{sec:futurework:verificationtoolsolvercodesign}). More fundamentally,
every query comes from a project that was already verified with Z3, so each
query is, by construction, one that Z3 solves; a query Z3 could not solve would
have been rewritten by the developer before the project was committed.

\subsection{RQ2: Dafny and \fstar}\label{subsec:eval-mariposa}

\begin{figure}[t]
\centering
\begin{subfigure}[t]{0.49\textwidth}
\includegraphics[width=\textwidth]{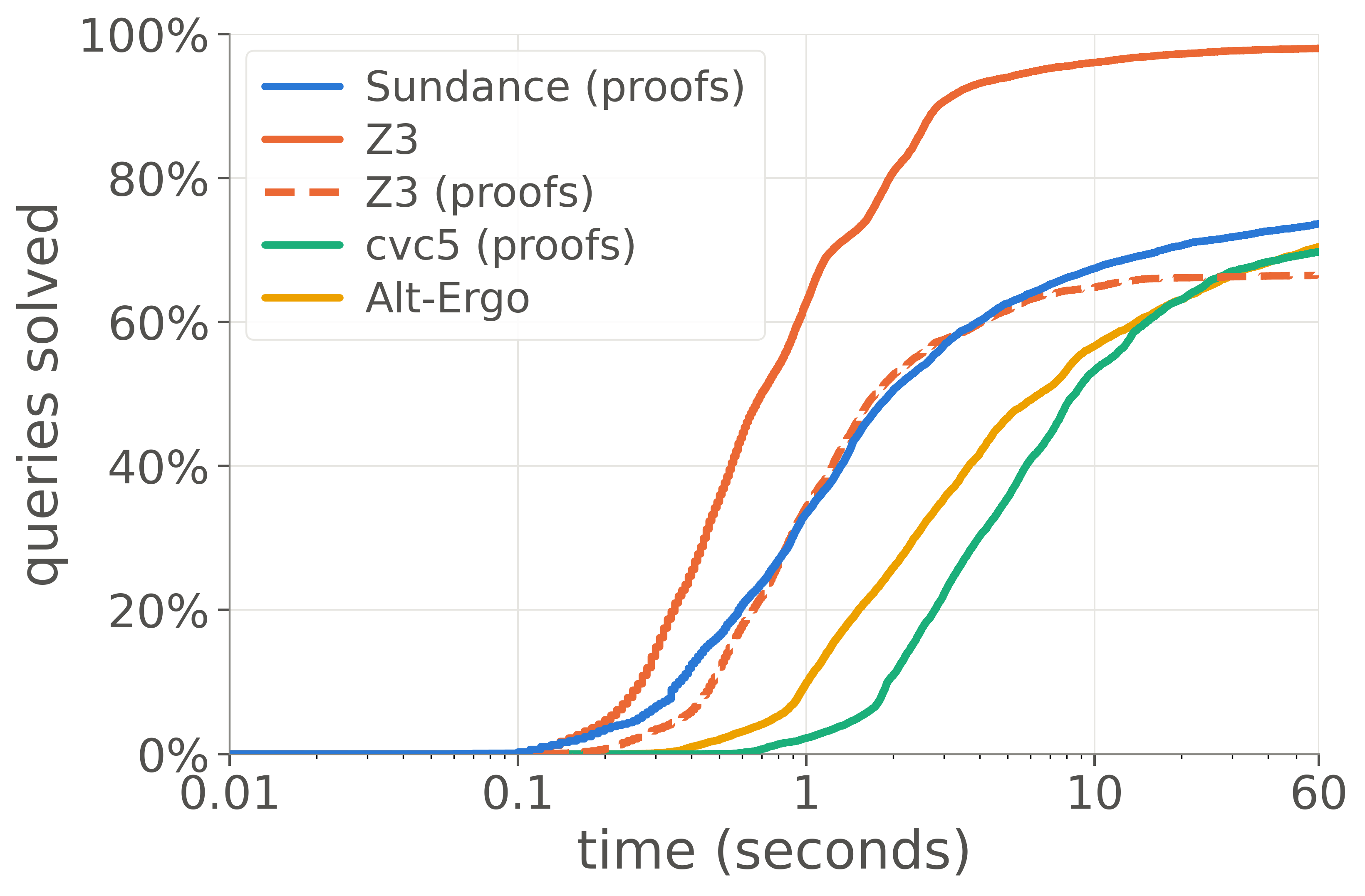}
\subcaption{Dafny.}\label{fig:eval-dafny}
\end{subfigure}%
\hfill
\begin{subfigure}[t]{0.49\textwidth}
\includegraphics[width=\textwidth]{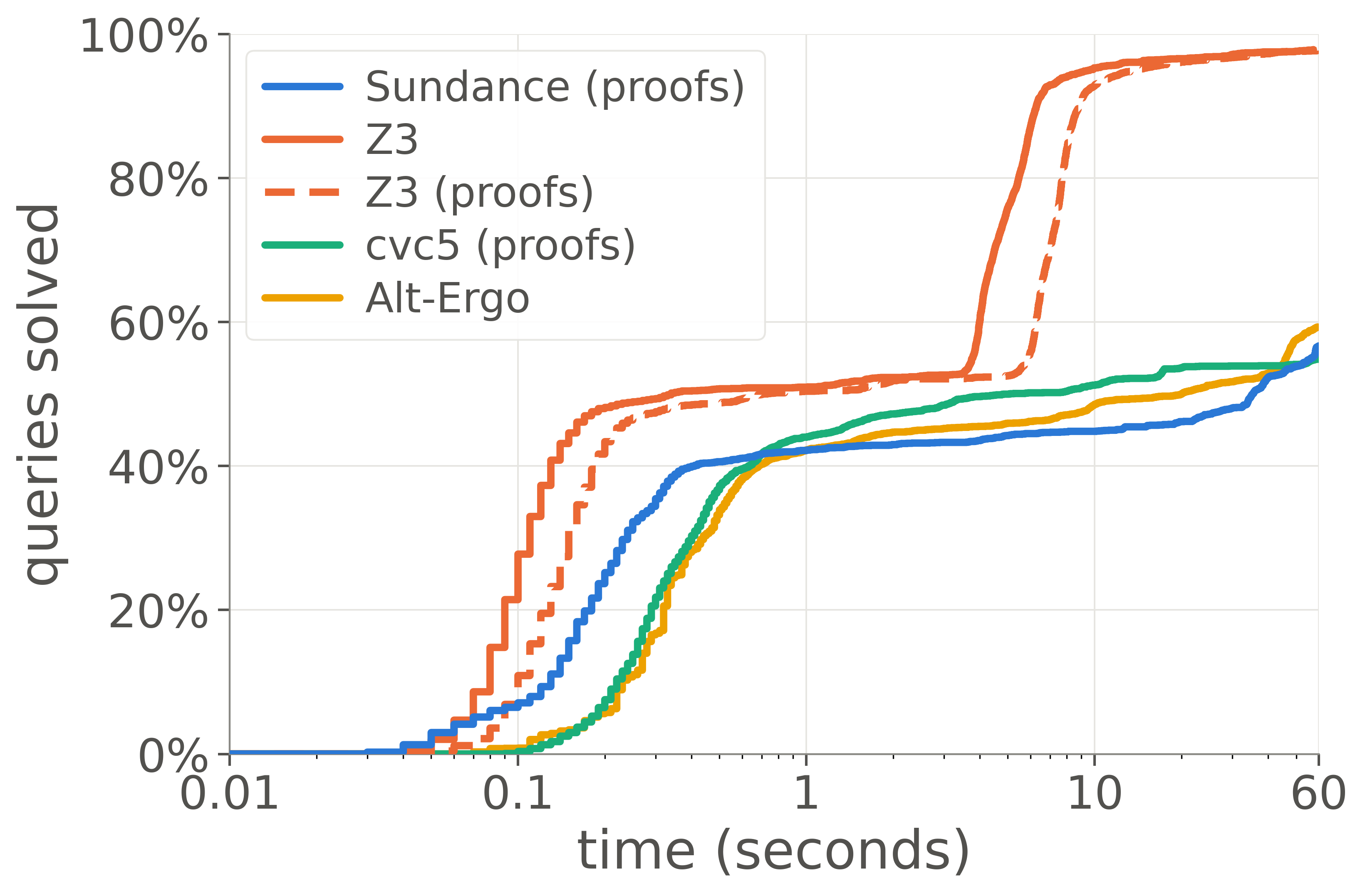}
\subcaption{\fstar.}\label{fig:eval-fstar}
\end{subfigure}
\caption{Percentage of Dafny and \fstar queries from \mariposa solved within a
given time budget. \toolname, Z3 (proofs), and cvc5 (proofs) produce proofs; Z3
and Alt-Ergo do not.}
\label{fig:eval-mariposa}
\end{figure}

On Dafny (\autoref{fig:eval-dafny}), \toolname solves 73.7\% of the queries
within 60\,s, more than cvc5 (69.8\%), Alt-Ergo (70.4\%), and Z3 with proofs
(66.5\%), while Z3 without proofs solves 98.0\%. 

The \fstar results (\autoref{fig:eval-fstar}) have a distinctive elbow. Most
solvers solve 40--50\% of the queries within about a second and a few more after
that, while Z3 solves most of the rest after about 4\,s. These more challenging queries
come from DICE$^\star$, whose queries are very large (a median of 14\,MB) and
take several seconds just to parse and preprocess. On DICE$^\star$, \toolname
solves 25.6\% of the queries within 60\,s, more than cvc5 (9.0\%) and Alt-Ergo
(22.7\%), whereas Z3 solves 96.8\%. On the other \fstar project, vWasm, the
solvers are again comparable (\toolname 84.8\%, Alt-Ergo 92.5\%, cvc5 96.4\%,
Z3 98.8\%). Overall, \toolname solves 56.7\% of the \fstar queries, between
cvc5 (54.9\%) and Alt-Ergo (59.3\%).

Unlike on Verus and Dafny, enabling proofs barely affects Z3 on \fstar (97.8\%
without proofs, 97.7\% with). We believe Z3's advantage on these queries comes
from its e-matching techniques~\cite{e-matching}, which keep matching efficient
over the very large sets of terms in DICE$^\star$; we discuss adopting them in
\autoref{sec:futurework}.

\subsection{RQ3: Varying Modules}\label{subsec:eval-modules}

\begin{figure}[t]
\centering
\begin{subfigure}[t]{0.49\textwidth}
\includegraphics[width=\textwidth]{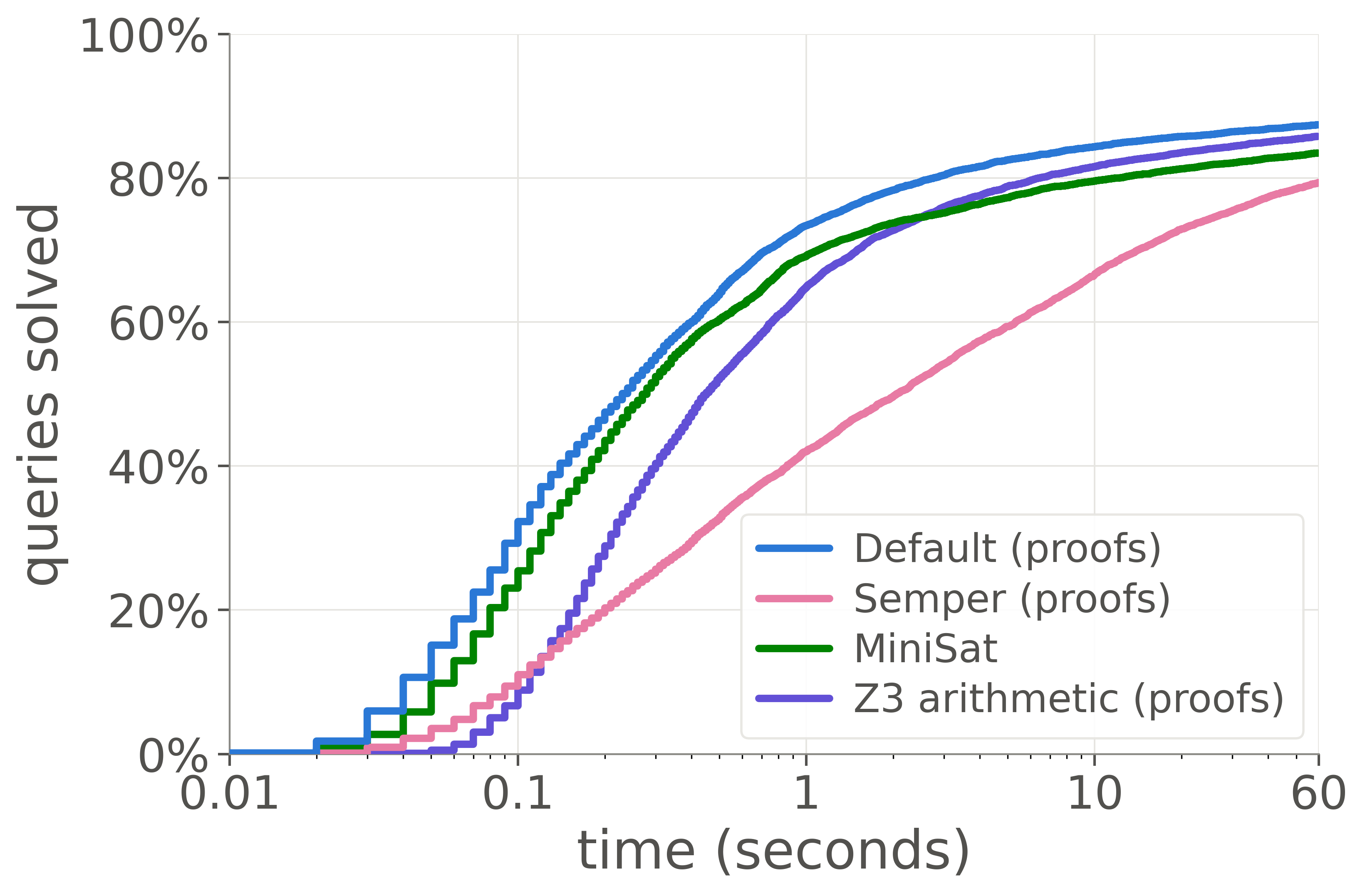}
\subcaption{Verus (development and held-out).}\label{fig:eval-modules-verus}
\end{subfigure}%
\hfill
\begin{subfigure}[t]{0.49\textwidth}
\includegraphics[width=\textwidth]{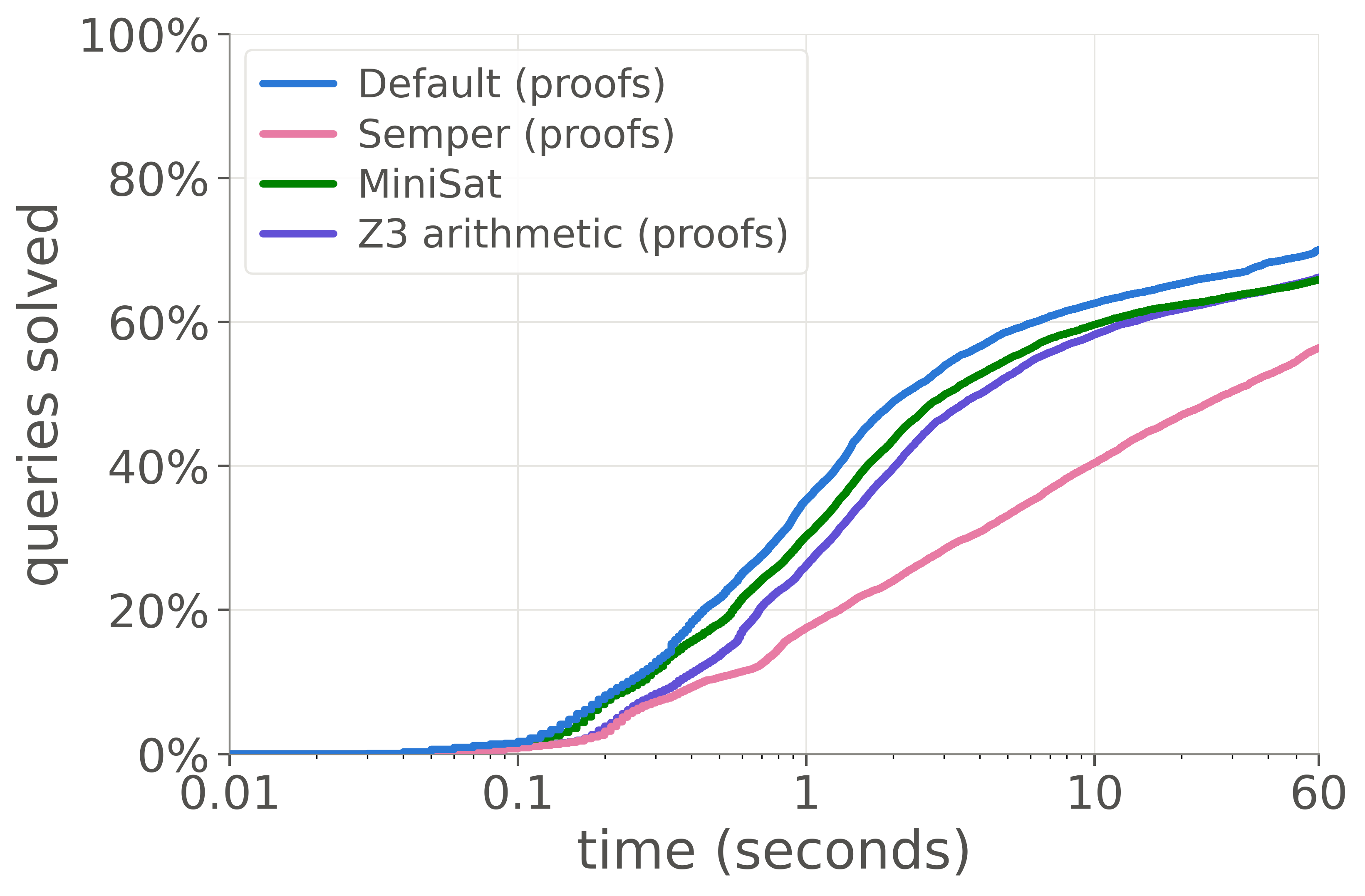}
\subcaption{Dafny and \fstar.}\label{fig:eval-modules-mariposa}
\end{subfigure}
\caption{Percentage of queries solved within a given time budget by \toolname's
default configuration and by variants that replace one module: the Semper
e-graph, the MiniSat SAT solver, or Z3's arithmetic solver.}
\label{fig:eval-modules}
\end{figure}

To evaluate \toolname's modularity, we replace one module at a time: the
e-graph with Semper~\cite{semper}, the SAT solver with
MiniSat~\cite{minisat} (\autoref{app:sat}), and the arithmetic solver with Z3's.
Each swap required no changes to the rest of the solver.
\autoref{fig:eval-modules} shows the results. Every variant solves fewer
queries than the default configuration on every benchmark set. Within 60\,s,
the default configuration solves 87.4\% of the Verus queries and 70.0\% of the
Dafny and \fstar queries, compared with 85.8\% and 66.2\% with Z3's arithmetic
solver, 83.5\% and 65.9\% with MiniSat, and 79.4\% and 56.4\% with Semper.

Each variant is slower for a different reason. MiniSat is a much simpler SAT
solver than \cadical and lacks many of its modern techniques. Semper is designed for
equality saturation workloads, rather than for the incremental, backtracking
use of an e-graph in an SMT solver. Semper's restore deletes the nodes created since
the snapshot, whereas \toolname expects terms to persist across backtracking,
so the adapter
re-inserts every term registered above the restore point after each restore
(\autoref{sec:egraph}). Profiling attributes 56\% of a quantifier instantiation
round to this re-insertion. 
Z3's arithmetic solver is fast on its own,
but its API does not match \toolname's incremental use: keeping it synchronized
with the SAT solver's assignments and backtracking requires redundant work that
\toolname's own solver avoids.

\section{Related Work}
\label{sec:related}

\subsubsection{SMT Solving.}\label{sec:related:smt}

Early program verifiers such as ESC/Java~\cite{esc-java} used
verification-specific provers like
Simplify~\cite{simplify}. Simplify's e-graph and pattern-based
quantifier instantiation~\cite{nelson-thesis} became the model for
later SMT solvers and inspired \toolname's matching
algorithm. Verifiers later moved to general-purpose SMT solvers,
largely for their performance.  \toolname revisits Simplify's focus on
verification while building on modern components such as CDCL SAT
solving and efficient proof formats. Alt-Ergo~\cite{altergo} is a
modern SMT solver designed for program verification and is used by
Why3~\cite{why3} and Frama-C~\cite{frama-c}; we compare against it in
\autoref{sec:eval}. SMTInterpol~\cite{smtinterpol} supports the logics
and features used by Verus, but its main focus is on interpolant
generation; it solves few of our Verus queries (\autoref{sec:eval}).
Other SMT solvers~\cite{yices2,opensmt2,mathsat5,bitwuzla,smtrat} lack
features required by program verification.

\subsubsection{Proof Production.}\label{sec:related:proofs}

Several SMT solvers can produce proof certificates that allow
unsatisfiability results to be checked independently. Z3 produces
proofs in a natural-deduction style format~\cite{z3-proofs} that is
specific to Z3, coarse-grained, and does not cover all of its theories
and features. cvc5 produces proofs in several formats, including the
cooperating proof calculus (CPC)~\cite{cvc5-proof} and
Alethe~\cite{alethe}. CPC is formalized in the Eunoia logical
framework and checked by Ethos~\cite{ethos} (written in C++) or by
Logos~\cite{logos} (written in Lean). Alethe proofs can be checked
with Carcara~\cite{alethe-checker}, but Alethe does not support as
many theories as CPC. In general, proof production adds overhead, and
may require some optimizations to be disabled. As a result, proofs are
rarely checked in practice: program verifiers do not check solver
proofs. \toolname instead uses eDRAT~\cite{edrat}, which extends the
clausal DRAT format used by SAT solvers~\cite{drat}. A clausal proof
is cheap to produce, and its propositional part can be checked with
existing SAT proof tools. We describe \toolname's proof production in
\autoref{sec:proofs}. A limitation of eDRAT is its inability
to capture preprocessing and simplification rules, but \toolname does
not rely on heavy preprocessing, only on conversion to CNF.

\subsubsection{Understanding and Debugging Solver Behavior.}\label{sec:related:debugging}

Because general-purpose solvers are opaque, a range of tools help users
understand why a query fails or behaves unpredictably. The Axiom
Profiler~\cite{axiom-profiler} and SMTScope~\cite{smt-scope} analyze Z3's
quantifier instantiation traces to expose matching loops and expensive axioms.
Verifier-level guides and tools help users diagnose and repair failing
proofs~\cite{dafny-debugging,fstar-debugging,proof-plumber}, and
Cazamariposas~\cite{cazamariposas} automatically locates the sources of
instability~\cite{mariposa} in a query. These tools work around the solver: they
reconstruct its behavior from traces tied to Z3's internals, which change across
versions. \toolname instead aims to be transparent by design, with few
heuristics and a small codebase that users can inspect directly.

\section{Future Work}
\label{sec:futurework}

We outline
several major avenues for future research and discuss how \toolname
can facilitate them.

\subsubsection{Instability.}\label{sec:futurework:instability}

A major challenge in SMT solvers is instability, where semantically
irrelevant changes in the input can lead to large variations in solver
performance~\cite{mariposa}. This is especially frustrating in
verification, where a developer's proof may fail because of a slight
change in the input, such as renaming a variable. Prior work mitigates
instability by normalizing queries~\cite{amrollahi-fmcad25,cav-normalization},
pruning irrelevant assertions~\cite{shake,dfyprune}, choosing patterns
or restructuring
proofs~\cite{butterfly,cutlerimproving,dafny64,dafny-stable-module},
or debugging unstable queries after the
fact~\cite{axiom-profiler,cazamariposas,instability-conjecture}.  These techniques help, but
none fully solves the problem, and all of them treat the solver as a
black box. Recent work has shown that stability can also be improved
inside the solver, by modifying Z3's quantifier instantiation
heuristics~\cite{cebeci2026}. In a large solver, however, such changes
must contend with many interacting components.  Because \toolname is
small and relies on few heuristics, it offers a platform for
controlling the sources of instability directly.

\subsubsection{Quantifier Heuristics.}\label{sec:futurework:quantifierheuristics}

Closely related to instability are the heuristics that govern
quantifier instantiation~\cite{cebeci2026}. Beyond e-matching, Z3
relies on several heuristics that decide which instances are created,
when they are created, and how long they are kept. We plan to explore
Z3 heuristics such as relevancy
propagation~\cite{z3-relevancy-propagation} and eager instantiation in
\toolname. In addition, Z3 uses garbage collection to discard terms
when the solver backtracks. We have not yet implemented such garbage
collection in \toolname, but we plan to do so in the future.

\subsubsection{Tailoring Solvers to Verifiers.}\label{sec:futurework:verificationtoolsolvercodesign}

Verifiers and their solvers are tightly coupled, but only in one direction:
verifiers tune their encodings, patterns, and options to Z3's heuristics, while
Z3 knows nothing about the verifier. This coupling is fragile; for instance,
Verus's upgrade from Z3~4.12.5 to 4.16.0 broke proofs in its standard library,
its tests, and six downstream projects~\cite{verus-z3-upgrade}. With a small,
modular solver, a verifier could control its own version of \toolname, control
when it changes, and adapt the solver to its encoding: for example,
instantiation heuristics aware of Verus's encoding. We see \toolname as a step
towards co-designing verifiers and their solvers.
 
\section{Conclusion}

\toolname is a new open-source SMT solver specialized for program
verification.  For reliability, \toolname is written in a safe
programming language and it is designed to be proof-producing from the
start. One of our goals is for \toolname to stay simple and modular,
and achieve comparable performance to state-of-the-art solvers on our
target applications. We hope \toolname will pave the way for
more research in SMT solving, and help researchers and
practitioners understand and control automation in verification tools.

\begin{credits}
\subsubsection{Disclosure of AI Use.}

We discuss our use of AI in \refappendix{app:ai}.

\subsubsection{Data Availability.}
\toolname is open source at \url{https://github.com/yaspar-org/Sundance-SMT}.

\subsubsection{Acknowledgments.}
We would like to thank Clark Barrett, Zachary Battleman, Armin Biere,
Nikolaj Bj\o{}rner, Martin Nyx Brain, Cayden Codel, Katalin Fazekas, Robert Jones,
Rustan Leino, Florian Pollitt, Long Qian, Giles Reger, Andy Reynolds, and
Max Willsey for insightful discussions and feedback during the development of Sundance. We would 
also like to thank Federico Mora for providing feedback on an earlier draft of this paper.

This work was supported in part by funding from AFRL and DARPA under Agreement
FA8750-24-9-1000, the Future Enterprise Security initiative at Carnegie
Mellon CyLab (FutureEnterprise@CyLab), and support from the Beneficial AI Foundation.

\end{credits}

\newpage

\bibliographystyle{splncs04}
\bibliography{refs}

\appendix
\maybeappendix{Use of AI}
\label{app:ai}

We began working on this project in May 2025. At that time, we found AI tools to
not be significantly helpful except for asking questions and generating small,
boilerplate code. The core of \toolname was mostly developed by the end of 2025
in this manner with a very small amount of AI-generated code. We open-sourced
\toolname shortly thereafter at the start of 2026. After open-sourcing the code,
all changes to \toolname were done via a pull request to the GitHub repository
and required at least one human review before being merged. Starting around
March 2026, we found AI tools to be significantly more helpful for generating
code, refactoring, and providing suggestions for improvements. As more powerful
models became available, the scope of AI assistance in our development process
expanded. For instance, some optimizations to the incremental arithmetic solver
were found with the assistance of AI tools. The
role of AI tools has surprised us in its rapid evolution and increasing impact
on our development workflow. We believe AI tools will continue to play a
significant role in the future development of SMT solvers.

We have used AI tools to assist in brainstorming, creating figures, drafting,
and editing the text of the paper. Any text generated by AI tools was reviewed
and edited by the authors both for style and accuracy.

\maybeappendix{Swapping the SAT Solver}
\label{app:sat}

\begin{figure}[t]
\centering
\includegraphics[width=\textwidth]{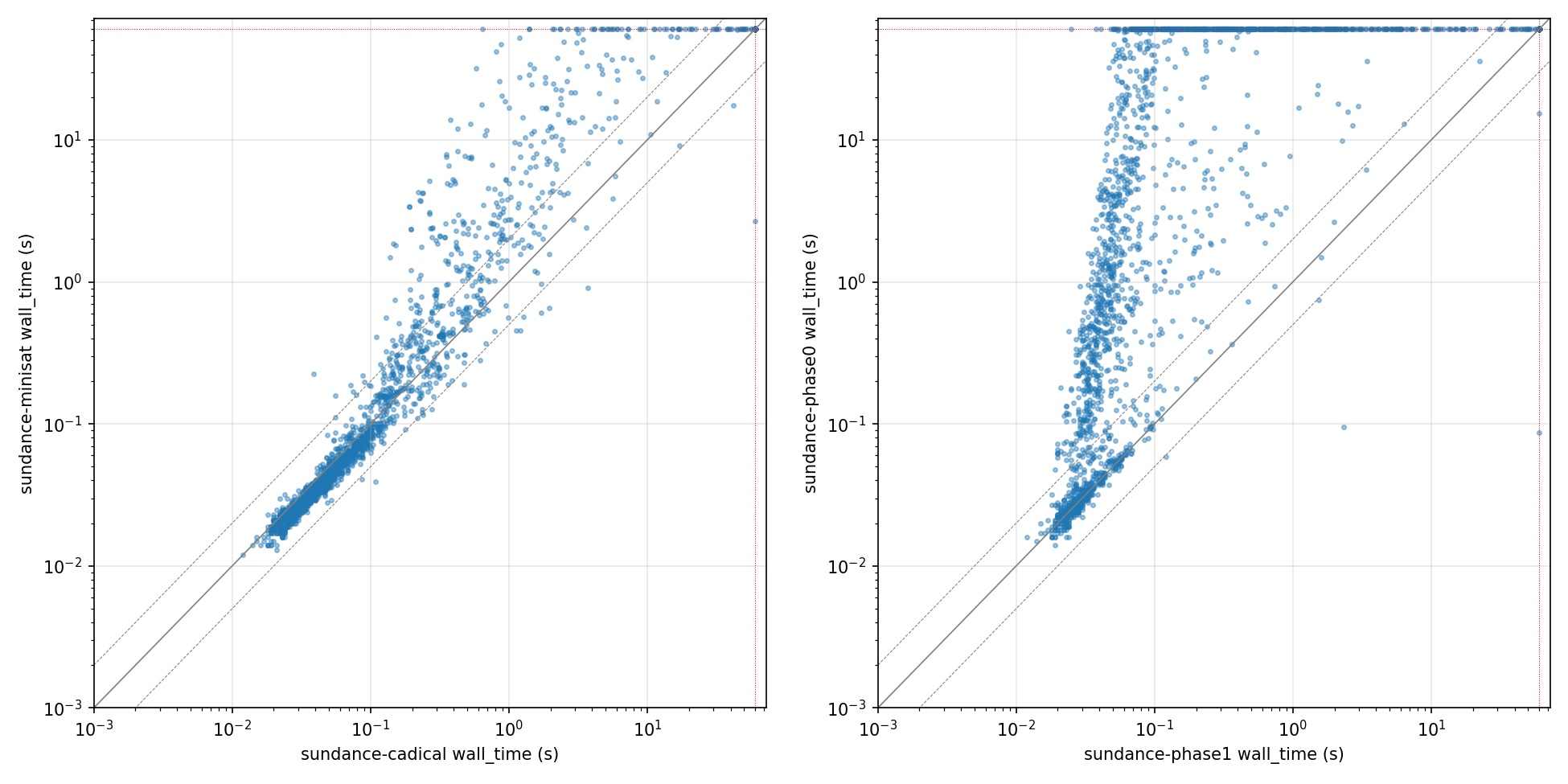}
\caption{
Per-query solving time on Verus queries, in seconds, with Z3's arithmetic
solver as the arithmetic backend; points above the diagonal
are slower on the vertical axis.
\emph{Left:} \toolname with \cadical ($x$-axis) vs.\ \toolname with MiniSat
($y$-axis).
\emph{Right:} an experiment using \cadical in both configurations,
with its initial variable phase set to its default of true
(\texttt{phase1}, $x$-axis) vs.\ set to false (\texttt{phase0},
$y$-axis), showing that choice of initial phase affects performance.
}
\label{fig:eval-minisat}
\end{figure}

As described in \autoref{sec:satsolver}, \toolname supports MiniSat~\cite{minisat} (via MiniSatUP~\cite{minisatup-github}) as an alternative to \cadical~\cite{cadical}, with MiniSat's initial variable phase set to true to match \cadical's default rather than MiniSat's default of false.
\autoref{fig:eval-minisat} compares the two SAT solvers on the Verus development set (\autoref{subsec:eval-verus}), using Z3's arithmetic solver as the arithmetic backend in both configurations. This differs from \autoref{sec:eval}, which uses \toolname's own arithmetic solver except in the Z3-arithmetic variant of \autoref{subsec:eval-modules}, so the numbers below are not directly comparable with those in \autoref{sec:eval}.
The results show that \toolname solves slightly fewer Verus queries with MiniSat than with \cadical as the backend SAT solver: 2,138 (83.0\%) with \cadical and 2,070 (80.4\%) with MiniSat, out of 2,576; 69 are solved only with \cadical and one only with MiniSat.
The right panel of \autoref{fig:eval-minisat} isolates the effect of initial phase by running \cadical with different phase settings, showing that the false phase alone accounts for a large slowdown and additional timeouts: with the false phase, \toolname solves only 1,436 queries (55.7\%).

\end{document}